\documentclass[journal]{IEEEtran}
\ifCLASSINFOpdf
\else
\fi
\usepackage{url}
\usepackage{graphicx}
\usepackage{amsmath}
\usepackage{tabularx}
\usepackage{multirow}
\usepackage{rotating}
\begin{document}
\title{Augmented Reality Warnings in Roadway Work Zones: Evaluating the Effect of Modality on Worker Reaction Times}

\author{Sepehr~Sabeti, %~\IEEEmembership{William States Lee College of Engineering, University of North Carolina at Charlotte,}
        Fatemeh~Banani~Ardecani, %~\IEEEmembership{William States Lee College of Engineering, University of North Carolina at Charlotte,}
        and~Omidreza~Shoghli %~\IEEEmembership{William States Lee College of Engineering, University of North Carolina at Charlotte}% <-this % stops a space
\thanks{Corresponding Author: Omidreza Shoghli, oshoghli@charlotte.edu}}% <-this % stops a space

\maketitle

\begin{abstract}
Given the aging highway infrastructure requiring extensive rebuilding and enhancements, and the consequent rise in the number of work zones, there is an urgent need to develop advanced safety systems to protect workers. While Augmented Reality (AR) holds significant potential for delivering warnings to workers,  its integration into roadway work zones remains relatively unexplored. The primary objective of this study is to improve safety measures within roadway work zones by conducting an extensive analysis of how different combinations of multimodal AR warnings influence the reaction times of workers. This paper addresses this gap through a series of experiments that aim to replicate the distinctive conditions of roadway work zones, both in real-world and virtual reality environments. Our approach comprises three key components: an advanced AR system prototype, a VR simulation of AR functionality within the work zone environment, and the Wizard of Oz technique to synchronize user experiences across experiments. To assess reaction times, we leverage both the simple reaction time (SRT) technique and an innovative vision-based metric that utilizes real-time pose estimation. By conducting five experiments in controlled outdoor work zones and indoor VR settings, our study provides valuable information on how various multimodal AR warnings impact workers reaction times. Furthermore, our findings reveal the disparities in reaction times between VR simulations and real-world scenarios, thereby gauging VR's capability to mirror the dynamics of roadway work zones. Furthermore, our results substantiate the potential and reliability of vision-based reaction time measurements. These insights resonate well with those derived using the SRT technique, underscoring the viability of this approach for tangible real-world uses.
\end{abstract}


% \fi
\IEEEpeerreviewmaketitle

\section{Introduction}
\IEEEPARstart{R}{oadway} work zones are pivotal in inspecting, maintaining, and upgrading transportation infrastructure to ensure their optimal function. Despite their essential role, these work zones carry a considerable degree of safety risks for the workers involved. These hazards often rise due to the proximity to moving vehicles, night work, extended work hours, and exposure to extreme weather elements  \cite{al2020does,hou2020study}. An examination of workplace fatalities and injuries around the world underscores the severity of the hazardous environment in roadway work zones. From 2003 to 2017, the CDC reported \cite{CDC} 4,444 deaths at U.S. road construction sites, averaging 123 fatalities annually. In 2022, workers struck by moving vehicles in Great Britain was ranked as the second most common cause of fatal accidents \cite{unite2022}. 

Recently, there has been a significant increase in investments aimed at building efficient transport infrastructure. These financial commitments are shaped by initiatives such as the Infrastructure and Investment Jobs Act \cite{housebill3684} in the United States and multibillion-dollar projects under strategic plans for sustainable, safe, and efficient transport infrastructure in the European Union \cite{eu2023}. Such actions are expected to cause a considerable increase in the number of roadway work zones in various countries. Given the upcoming increase in the number of work zones, it is crucial to prioritize improving the safety systems within roadway work zones.

Reflecting this need, recent trends within the work zone community \cite{nnaji2020improving,sakhakarmi2021tactile,national2022use} have increasingly emphasized the role of advanced technology in mitigating safety risks. However, most of the available technologies are primarily reactive \cite{tapco_sonoblaster,highway_resource_intellicone}, with shortcomings in the effective delivery of warnings. The primary challenge lies in the efficacy of universal warning mechanisms employed by distance-reaching technologies, such as loud audio-based sirens, which tend to lose their impact amid the noise of roadway work zones. As a result, researchers have shifted their focus towards developing mechanisms that deliver warnings in closer proximity to workers through individual sensor placement. However, most of these technologies have predominantly emphasized the use of wearable haptic sensors alone \cite{wu2022real, sakhakarmi2021tactile}. Consequently, there is limited information available regarding the impact on reaction time when employing warning delivery methods that leverage visual cues, other sensory modalities, or their combinations within the context of roadway work zones.

In the meantime, notable progress in both hardware and software for Augmented Reality (AR) technology has led to considerable enhancements in its effectiveness for delivering warnings within the transportation and construction sector\cite{ramos2022proposal,li2018critical,gilson2020leveraging}, presenting promising horizons. Yet, when it comes to roadway work zones, the use of AR technologies is not sufficiently studied. Previous research studies \cite{calvi2020effectiveness,matviienko2022scootar,von2020augmentation}, although extensive across various applications, have inadequately addressed the impacts of multimodal designs of AR warnings on worker reaction times, specifically within roadway work zones. The design of safety systems for roadway work zones demands more than just general reaction time assumptions. The distinctive characteristics of these work zones, including the cognitive load associated with physical activities and the sensory taxing environment, can have a substantial impact on workers' reaction time performance. Therefore, it is essential to consider these factors when developing time-critical safety warnings, taking into account the unique challenges faced by workers in roadway work zones.

Moreover, a major challenge in researching intrusion alert technologies within roadway work zones is the inherent safety risks and rarity of intrusion events in real-world settings. Consequently, conducting high-fidelity studies to assess reaction times in these contexts is not only complex and resource-intensive, but also carries significant risks \cite{zhang2019crash, awolusi2019active}. To this end, Virtual Reality (VR) has emerged as a practical alternative to replicate scenarios that are costly and logistically challenging. In the context of roadway work zone safety, a few studies \cite{li2023semi,ergan2022developing,zou2020integrated} have utilized VR to explore worker behavior across different contexts and technological settings. Also, some studies within healthcare and military sectors \cite{horikawa2022comparing,zaman2021investigating,merenda2019effects} have reported the use of AR simulations within VR environments. However, a critical question is the extent to which VR-simulated AR can effectively mimic the complexities of real-world dynamics and accurately record reaction times in the context of safety within roadway work zones. This consideration is also vital for future simulation-based training applications, as it has a direct impact on the reliability and validity of using VR simulations to train workers for real-world scenarios.

To address these challenges, we devised a comprehensive approach aimed at measuring the reaction times to different multimodal AR warnings in roadway work zones. Leveraging the AR-enabled innovative warning technology, developed by our team \cite{sabeti2021toward}, we designed a methodology grounded in three key components: an advanced AR system prototype, virtual reality simulation of the AR system's functionality as well as the roadway work zone environment, and the Wizard of OZ methodology for synchronizing user experiences across experiments. We used a hybrid approach combining between-subject and within-subject designs to study the effects of varying conditions on reaction times and the effectiveness of AR-based warnings. This involved conducting five experiments in both outdoor controlled work zone and indoor VR settings, enabling us to extensively capture data on multimodal AR warnings' performance. The primary research questions addressed in our experimental design are as follows:

\begin{itemize}
    \item \textbf{RQ1.} Which multimodal augmented reality warning configuration is the most effective in increasing the safety of workers in roadway work zones, as measured by the reaction time to safety risks?
    \item \textbf{RQ2.} Is there a statistically significant relationship between the reaction times to simulated multimodal AR warnings in a virtual reality environment of the roadway work zone, and the reaction times to actual AR warnings in real-world outdoor settings?
    \item \textbf{RQ3.} Can the utilization of vision-based pose tracking effectively detect reaction patterns and serve as an indicator for reaction times to AR warnings?

\end{itemize}

This study contributes towards the enhancement of roadway work zone safety by offering a comprehensive set of reaction time metrics and benchmarks that are specifically designed for the unique environment of roadway work zones. These benchmarks are particularly relevant to the design of multimodal warning delivery systems, serving as a reference for developing real-time safety systems that leverage AR technology. This furthers our understanding of how to best utilize AR and multimodal warnings to enhance worker safety by reducing the reaction times of workers. Furthermore, we extend existing knowledge by exploring the use of VR as a simulation tool for AR-based warning systems, specifically in the context of safety research around reaction times. By drawing comparisons between the reaction time in VR-simulated AR and real-world AR scenarios, we enhance our understanding of VR simulation's effectiveness in reaction time measurement. Additionally, this study introduces the use of vision-based pose tracking to assess workers' reaction times, enhancing the overall understanding of occupational safety within roadway work zones. By tracking body movements in real-time, it's possible to evaluate how quickly workers respond to safety alerts. 
This non-intrusive system would monitor a worker's normal operational movements, and upon issuing a safety alert, compare the latency period between the alert and the detected change in the worker's movement pattern. Over time, these data can help determine average reaction times, identify individuals or situations where reaction times are slower than expected, and develop interventions to improve response rates. This could also assist in creating personalized training programs for workers with disabilities. This approach not only enhances the safety and well-being of workers, but also contributes to the overall efficiency and safety of roadway work zones.

\section{Related Work}

%Reviewing previous work, we first evaluate existing safety measures in roadway work zones, then investigate trends in warnings. Next, we investigate the methods of measuring reaction times and their implications in the context of highway work zone safety. Finally, we explore the role of virtual reality in simulating augmented reality across multiple fields.

\subsection{Safety Measures in Roadway Work Zones}
Despite acknowledging the risks associated with roadway work zones in various studies, the implementation of new technologies to improve safety in these environments has been limited. Existing safety measures primarily rely on reactive approaches \cite{gambatese2017work, park2019embedded}. These mechanisms include portable signs, automated flaggers, directional alarms, and warning lights. On the contrary, the dynamic environment of highway work zones accentuates the need for more robust safety systems that can quickly adapt to the changes in the working environment of highway workers. There is a growing shift toward smarter and more proactive safety systems in roadway work zones. A systematic review conducted by Nnaji et al. \cite{nnaji2020improving} highlighted the growing need for adopting smart automated technologies and a departure from traditional approaches, a shift further propelled by the emergence of advanced sensing technologies.

The roots of modern work zone intrusion technologies can be traced back to 1993, when Stout et al. \cite{stout1993maintenance} introduced an early design of what became the foundation of modern intrusion systems as part of the Strategic Highway Research Program. This program pioneered wireless and pneumatic sensor-based systems for work zones. Although these initial systems faced challenges in terms of adoption, they served as the basis for future advancements in the field and inspired the development of radar-based technology, allowing for vehicle speed and trajectory tracking \cite{eseonu2018reducing}. Another category of work zone intrusion technologies relies on sensor-based mechanisms that detect intrusions by mechanical impact \cite{marks2017active}. These sensors are typically integrated into channelizing devices that define the boundaries of work zones. Different combinations of these approaches have also driven the evolution of effective and comprehensive work zone intrusion systems.

Although limited, there have been some efforts to develop advanced technologies with potential applications in roadway work zones. For instance, a recent research study by Sakhakarmi et al. \cite{sakhakarmi2021tactile} focused on a proximity-based alerting system that uses tactile signals as the primary mode of communication with users. However, the effectiveness of this system in work zones on the roads can be compromised by high levels of noise and cognitive demands placed on workers. In another study, Chan et al. \cite{chan2020incorporating} proposed a wearable-based hazard proximity warning system to enhance the awareness of construction workers. Although this system utilizes proximity-based triggering mechanisms, it shares the limitations of reactive systems that are only activated when hazards are in close proximity. Additionally, Kim et al. \cite{kim2023signal} proposed an Internet of Things (IoT)-based proximity warning system that notifies workers when they are in close proximity to heavy equipment. While these studies show promise, some challenges still remain, most notably potential malfunction in noisy and demanding nature of work environments. Future research is needed to overcome these limitations by incorporating multiple modes of communication and multimodal warning systems, integrating real-time data, and effectively addressing the unique challenges faced by workers in work zones.
\subsection{Warnings Modality and Design}
A fundamental question is the choice between unimodal and multimodal warning alternatives. Various researchers have attempted to address this question in different contexts, revealing that the efficacy of warning designs is heavily influenced by the specific context and environment. Several studies have evaluated the performance of different sensory modes of warnings in diverse scenarios. For example, Yun et al. \cite{yun2020multimodal} conducted human-in-the-loop experiments to compare various warning combinations, incorporating visual, auditory, and haptic modalities, during conditionally automated driving take-over request scenarios. Their findings showed that the visual-auditory-haptic modal combination exhibited the best performance in both human behavioral and physiological data, while visual-auditory warnings were most effective in vehicle data. The combination of visual-auditory-haptic warnings outperformed all other modes on various performance indices. 

On the contrary, the use of visual-only warnings, typically used in manual driving, resulted in the worst performance in conditionally automated driving situations. In another study, Matviienko et al. \cite{matviienko2018augmenting} explored the use of multimodal warning signals to increase the awareness of children cyclists and prompt action in critical situations. They developed a bicycle simulator equipped with these signals and found out that participants spent more time perceiving visual cues compared to auditory or vibrotactile cues. Unimodal signals were easier to recognize and suitable for conveying directional cues. However, when prompting stop actions, the reaction time was shorter when all three modalities were simultaneously used. Furthermore, Wang et al. \cite{wang2022effect} investigated the effects of visual and auditory icons on the effectiveness of multimodal audiovisual warnings. They found that multimodal warnings provided additional benefits to drivers: the incorporation of high mapping auditory signals into multimodal warnings led to improved driving performance. While there is no universal consensus on a definitive guideline for warning design, these studies suggest that multimodal warnings tend to outperform unimodal designs of warnings, particularly in time-sensitive contexts. The choice between unimodal and multimodal warnings should be carefully considered based on the specific requirements and demands of each situation to ensure optimal effectiveness of systems and safety.

Despite the abundance of literature on digital warnings in general, a significant gap exists in the research regarding warning designs and modality in AR environments in the context of real-time safety systems for Vulnerable Road Users (VRUs) \cite{lazaro2021interaction}. The majority of previous studies \cite{wiegand2019incarar, zhou2018arve, calvi2020effectiveness} have focused mainly on in-vehicle AR warnings, leaving a gap in understanding how to effectively design AR warnings for workers and other VRUs. However, recent studies have started to change this trend. For example, Matviienko et al. \cite{matviienko2022scootar} investigated the enhancement of the safety of electronic scooters with unimodal warnings, including AR warnings, vibrotactile feedback on the handlebar, and auditory signals, to prevent collisions with other road users. Similarly, Von et al. \cite{von2020augmentation} explored potential approaches for AR applications to enhance cyclist safety and conducted a pilot study. These studies have considered multimodal designs, which incorporate visual, audio, and haptic feedback in warnings. However, there is a research gap on the specific examination of the impacts of AR-based warnings on roadway workers and their reaction times in various settings and contexts. This is particularly true when it comes to the design of multimodal warnings.

\subsection{Reaction Time: Measures, Influences, and Applications to Roadway Worker Safety}
Reaction Time (RT) measurement has been an important tool in psychology and neuroscience for more than a century \cite{posner2005timing}. RT measurements have been widely used by researchers \cite{fernandez2011relation,li2004transformations} to explore a diverse range of cognitive processes, study perception, and examine how quickly individuals can detect and interpret sensory stimuli from their environment. To this end, the Simple Reaction Time (SRT) task is widely recognized as the most common method to measure reaction time \cite{willoughby2018benefits}. In this approach, participants are instructed to respond as rapidly as they can to a single stimulus. The stimulus can take various forms, including visual cues, auditory signals, or somatosensory stimuli. Participants typically execute their response by pressing buttons or keys, or performing vocal reactions \cite{mueckstein2022modality,maslovat2019effect}. The review of the literature suggests that reaction time can be influenced by various factors such as age \cite{richer2014impact}, sex \cite{lu2020increased, woods2015age}, attention \cite{greenwald2022sequential}, fatigue \cite{langner2010mental}, %arousal levels \cite{ulrich1996does},
and task complexity \cite{maylor1992effects}.

Meanwhile, understanding workers' reaction time to safety warnings plays a vital role in the development of effective alert technologies. This significance is particularly accentuated in the context of roadway work zones, where the complex environment and the presence of fast-moving vehicles require a timely and rapid response from workers in case of intrusions \cite{nnaji2020case}. To this end, several studies have been conducted to investigate workers' reaction time in various systems and working environments related to roadway work zones. For example, Thapa et al. \cite{thapa2021using} examined the optimal configuration of a work zone intrusion alert technology and explored the relationship between sensor placement and alerting modules, considering workers' naturalistic reactions. In another research work, Nnaji et al. \cite{nnaji2020case} provided guidelines for the adoption of different commercially available work zone technologies for roadway work zones, taking into account the workers' reaction time and response rate as essential metrics in their framework. In another study, Awolusi et al. \cite{awolusi2019active} quantified the reaction time of roadway workers to two commercially available intrusion alert technologies specifically designed for roadway work zones. Finally, in a recent study, Yang et al. \cite{yang2023vibrotactile} conducted three experiments to assess the viability of using vibrotactile signals as warnings for road workers. The experiments aimed to assess the perception and performance of the generated signals at different body locations and to examine the usability of various warning strategies. Our review suggests that the existing literature does not provide sufficient evidence to provide insights into reaction times specifically related to AR-based warnings in this particular field.

\subsection{Virtual Reality Simulations for Evaluation of AR Warnings}
The rapid development and widespread adoption of AR technology have propelled it to the forefront of various domains and industries, offering a multitude of applications and possibilities \cite{li2018critical, bottani2019augmented}. However, the accelerated growth of AR has created a demand for efficient methods to prototype and evaluate AR interfaces and interactions in a timely manner \cite{kelly2018arcadia, muller2021spatialproto}. Traditional approaches such as wire-framing and paper sketching, while useful in certain design contexts, often fall short of capturing the true essence of the user experience in AR. These methods struggle to convey the immersive and context-sensitive aspects of AR applications, making it challenging to evaluate user interactions and gather meaningful feedback on usability and functionality. To overcome these limitations, researchers and practitioners have turned to alternative approaches that take advantage of virtual environments, specifically virtual reality, for prototyping and evaluation of AR applications. Using VR simulations, designers and developers can create virtual representations of AR interfaces and interactions that closely mimic the real-world user experience. This immersive and interactive environment allows for more realistic user testing and evaluation, allowing stakeholders to gain a deeper understanding of the usability, effectiveness, and user satisfaction of AR applications. 

In order to achieve this specific objective, several studies have been conducted. For example, in the field of surgical applications, Hettig et al. \cite{hettig2018ar} used VR simulations to mimic AR environments and investigated individual parameters for surgical procedures. By simulating the AR experience, they were able to explore different scenarios and optimize the application of AR technology in this novel context. In another study, Terrier et al. \cite{terrier2018evaluation} focused on the impact of registration errors between virtual and real objects in AR. They used VR simulations to control experimental conditions and examine the effects of registration errors. In the realm of public safety applications, Grandi et al. \cite{grandi2021design} proposed a framework that used virtual reality to evaluate AR interfaces in traffic stops and firefighting search and rescue scenarios. Using virtual reality simulations, they were able to simulate realistic situations and gather feedback to improve the design and effectiveness of AR interfaces in these critical public safety contexts. Furthermore, Zaman et al. \cite{zaman2021investigating} conducted a study focused on the usability of AR technology in combat missions. They used virtual reality simulations during subterranean operations to investigate the usability aspects of AR interfaces. Finally, Burova et al. \cite{burova2020utilizing} used VR AR simulation along with gaze tracking to evaluate the effectiveness of AR guidance and safety awareness features for elevator maintenance. Through an iterative development-evaluation process, industry experts participated in testing and providing feedback on the AR simulation and gaze tracking system. The study also included a survey that used actual gaze data from the evaluation to collect comments and insights from industry experts regarding the usefulness of the AR simulation and gaze-tracking approach. 
\section{Methodology}
\begin{figure}[t!]
    \centering
    \includegraphics[width=3.2in]{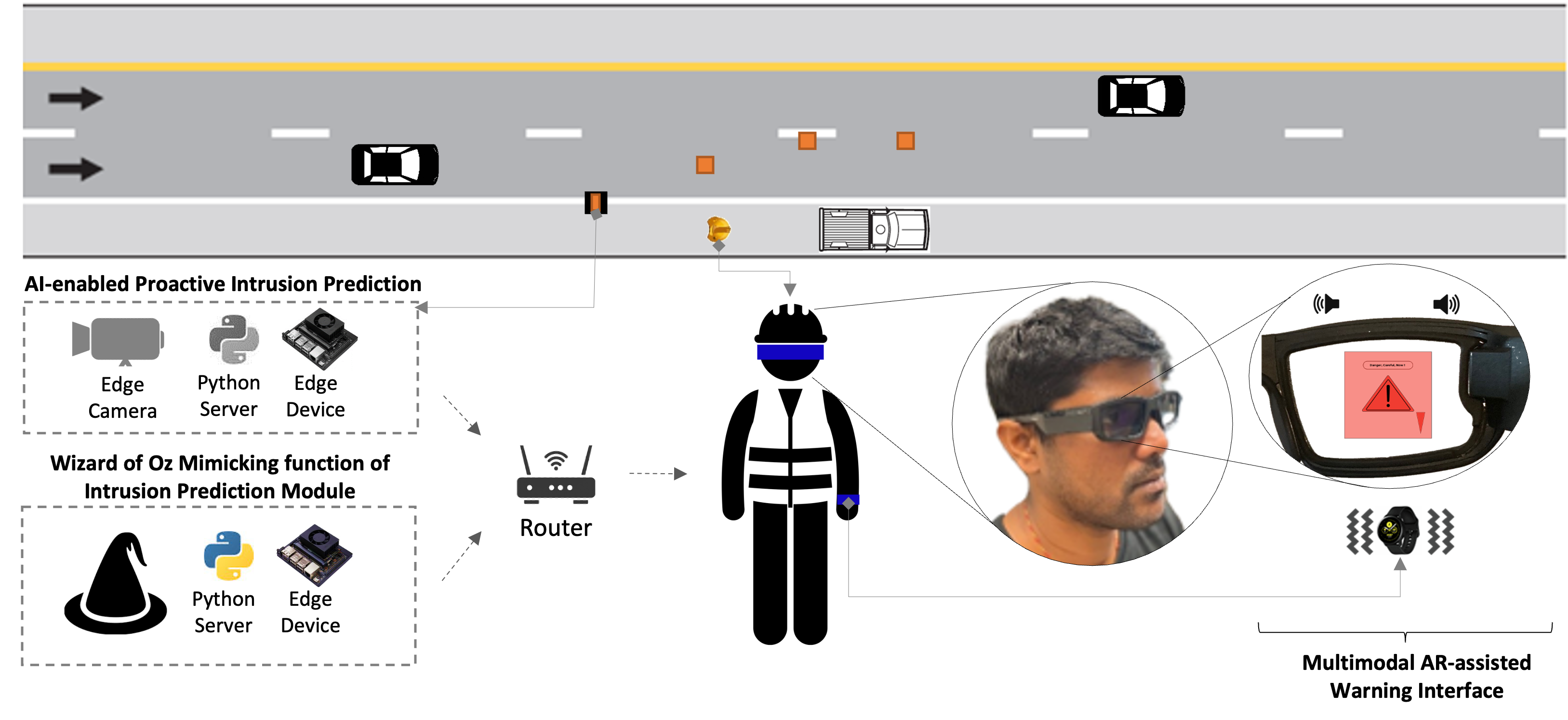}
    \caption{Overview of the Augmented Reality-Based Safety Technology and Its Warning Interface Features}
    \label{fig:overview}
\end{figure}
%\begin{figure*}[!t]
%\centering
%\subfloat[Case I]{\includegraphics[width=2.5in]{box}%
%\hfil
%\subfloat[Case II]{\includegraphics[width=2.5in]{box}%
%\label{fig_second_case}}
%\caption{Overview of the Augmented Reality-Based Safety Technology and Its Warning Interface Features.}
%\label{fig:overview}
%\end{figure*}

\subsection{Study Overview}
To achieve the objectives of this study, we designed a comprehensive approach to measure reaction times in distinctive setups to evaluate the efficacy of various sensory modes of warning deliveries as measured by reaction time. For this purpose, we conducted a desktop-based simple reaction time measurement that served as our baseline. The objective here was to create a standard benchmark against which we could compare the reaction times to different multimodal warnings observed in other scenarios. Next, we conducted a test in a controlled outdoor work zone setting. Here, participants were equipped with AR glasses that delivered warnings about potential intrusions or hazards. The aim was to evaluate the impact of different combinations of sensory modes of AR warnings on the reaction time of workers in a realistic environment. Finally, we conducted a series of experiments within an immersive virtual reality environment that mimicked a roadway work zone. The intent of this setup was twofold. Firstly, we sought to validate the fidelity of our VR work zone simulations by comparing the participants' reaction times in the VR environment with those from the real-world test. Second, we used this setup to measure reaction times using vision-based pose-tracking algorithms by capturing and analyzing the movements of the participants in response to the simulated warnings. In the following, we will detail our methodological framework, discuss the AR and VR technologies used, explain the design and execution of our experiments, and elaborate on other significant aspects integral to addressing the research questions. Table \ref{table:detail} also provides comprehensive information on the details, specifications, apparatus, and warning specifications of the experiments conducted in this study.

\subsection{Experimental Apparatus and Setup}
\subsubsection{Augmented Reality Technology}
This study utilized an innovative AR warning technology, conceptualized and developed by the authors' team, as detailed in our previous work \cite{sabeti2021toward,doi:10.1080/10803548.2023.2295660}. This framework, as shown in Figure \ref{fig:overview}, consists of two main components: the AI-powered back-end, which processes real-time video data to predict potential intrusions in the work zone, and the front-end, which uses a multimodal AR interface to alert workers in real time about immediate risks. This AR warning interface incorporates visual, auditory, and haptic cues and provides workers with timely warnings regarding possible intrusions or hazards, improving their situational awareness, and facilitating quick and appropriate responses to ensure worker safety. In this study, we used the Wizard of Oz methodology (WOZ) to replicate the functionality of the back-end component without using an actual AI module. We replaced the role of AI with a pre-defined scripted "wizard" written in Python that runs on the edge device. We adopted this approach to focus on specific intrusion scenarios that activate the warning interface within a controlled environment, as these occurrences are infrequent in real-world situations. Using this methodology, we could iteratively record participants' reaction times and responses to AR warnings. Additionally, the approach ensured that participants encountered similar and synchronized scenarios with consistent trigger stages across multiple experiments. The back and front ends were connected within a local network facilitated by a router.

\begin{figure}[ht!]
    %\centering
    \includegraphics[width=\linewidth]{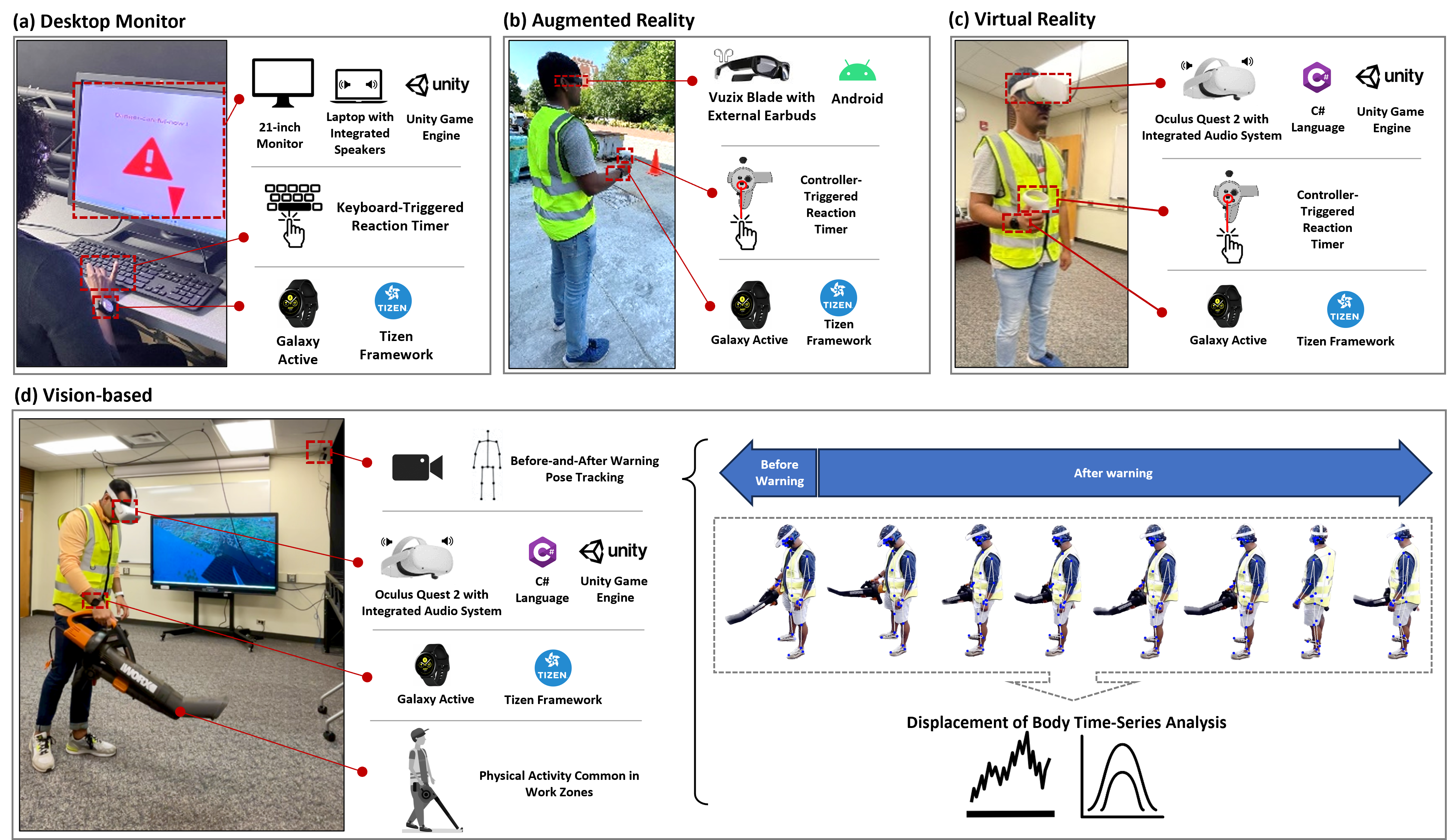}
    \caption{Hardware, Software, and Techniques Utilized in Different Approaches for Reaction Time Measurement. (a) A participant interacts with a keyboard, responding to a variety of stimuli (b) participant employing a handheld device to react to stimuli presented via augmented reality glasses. (c) Immersed in a virtual reality environment, a participant uses a controller to respond to delivered stimuli (d) In a virtual work zone environment, a participant operates a leaf blower while cameras capture the body movements.}
    \label{fig:general}
\end{figure}

%\newpage
\begin{table*}[t!]
%\begin{sidewaystable}
  \centering
  \caption{Details and Specifications of the Designed Warnings and Experiments}
  \label{table:detail}
  \footnotesize
  \renewcommand{\arraystretch}{1.5}
  \scalebox{0.77}{

\begin{tabular}{|l|l|cc|cc|cc|c|c|}
\hline
\multicolumn{1}{|c|}{\multirow{2}{*}{\textbf{Experiments}}} & \multicolumn{1}{c|}{\multirow{2}{*}{\textbf{Contextual Environment}}}                                                                                                                                          & \multicolumn{2}{c|}{\textbf{Visual (V) Stimulus}}                                                                                                                                                                                                               & \multicolumn{2}{c|}{\textbf{Audio (A) Stimulus}}                                                                                                                                                                                                                            & \multicolumn{2}{c|}{\textbf{Haptic (H) Stimulus}}                                                                                                                                                                                                                                                                                              & \multirow{2}{*}{\textbf{\begin{tabular}[c]{@{}c@{}}Warning \\      Modalities Tested\end{tabular}}}        & \multirow{2}{*}{\textbf{\begin{tabular}[c]{@{}c@{}}Reaction Time \\      Measurement\end{tabular}}}                                              \\ \cline{3-8}
\multicolumn{1}{|c|}{}                                      & \multicolumn{1}{c|}{}                                                                                                                                                                                          & \multicolumn{1}{c|}{\textbf{Delivery}}                                                                                                    & \textbf{Design}                                                                                                    & \multicolumn{1}{c|}{\textbf{Delivery}}                                                                                     & \textbf{Design}                                                                                                                                & \multicolumn{1}{c|}{\textbf{Delivery}}                                                                                                                                                                                      & \textbf{Design}                                                                                                 &                                                                                                            &                                                                                                                                                  \\ \hline
A   - Baseline                                              & -Indoor Desktop-based                                                                                                                                                                                          & \multicolumn{1}{c|}{Computer Monitor}                                                                                                     & \multirow{5}{*}{\begin{tabular}[c]{@{}c@{}}\\ \\ \\ \\ \\ \\ \\ \\ Warning \\      Sign as\\       visualized \\      in Figure \ref{fig:notification}\end{tabular}} & \multicolumn{1}{c|}{\begin{tabular}[c]{@{}c@{}}Computer\\      Speakers\end{tabular}}                                      & \multirow{5}{*}{\begin{tabular}[c]{@{}c@{}}\\ \\ \\ \\ \\ \\ \\ \\ -Stereo \\      High-pitched \\      Beep\\      -44100 Hz \\ Frequency \\      -0.2 milliseconds\end{tabular}} & \multicolumn{1}{c|}{\multirow{2}{*}{\begin{tabular}[c]{@{}c@{}}Right-hand\\      Galaxy\\      Smartwatch\end{tabular}}}                                                                                                    & \multirow{5}{*}{\begin{tabular}[c]{@{}c@{}}\\ \\ \\ \\ \\ \\ \\ \\ Haptic\\      Feedback\\      Through \\ Tizen \\      Native API\end{tabular}} & \multirow{4}{*}{\begin{tabular}[c]{@{}c@{}}\\ \\ \\ \\ Four \\      Combinations\\      (V, AV, HV, HAV)\end{tabular}} & \begin{tabular}[c]{@{}c@{}}-Keyboard Click\\      -Simple Reaction\\      Time\end{tabular}                                                         \\ \cline{1-3} \cline{5-5} \cline{10-10} 
B - AR                                                        & \begin{tabular}[c]{@{}l@{}}-Outdoor\\      -Controlled Roadway\\      Work Zone Testbed\\      -No Traffic\\      -Natural Ambient Noise\end{tabular}                                                          & \multicolumn{1}{c|}{\begin{tabular}[c]{@{}c@{}}-AR Glasses\\      Display on \\      Right Lens\\      -Semi-transparent\end{tabular}}      &                                                                                                                    & \multicolumn{1}{c|}{\begin{tabular}[c]{@{}c@{}}Bluetooth\\      Earbuds\end{tabular}}                                      &                                                                                                                                                & \multicolumn{1}{c|}{}                                                                                                                                                                                                       &                                                                                                                 &                                                                                                            & \begin{tabular}[c]{@{}c@{}}-VR Right-hand\\      Controller\\       "A" Button\\      -Simple Reaction\\      Time\end{tabular}                   \\ \cline{1-3} \cline{5-5} \cline{7-7} \cline{10-10} 
C - VR-WOT                                                   & \begin{tabular}[c]{@{}l@{}}-Indoor\\      -VR Simulated \\      Roadway Work Zone\\      -No Traffic\\      -No Ambient Noise\end{tabular}                                                                     & \multicolumn{1}{c|}{\multirow{3}{*}{\begin{tabular}[c]{@{}c@{}}\\ \\ \\ \\ \\-VR Environment \\      Mimicking AR\\      -Semi-transparent\end{tabular}}} &                                                                                                                    & \multicolumn{1}{c|}{\multirow{3}{*}{\begin{tabular}[c]{@{}c@{}}\\ \\ \\ Built-in VR \\      Headset\\       Speakers\end{tabular}}} &                                                                                                                                                & \multicolumn{1}{c|}{\multirow{3}{*}{\begin{tabular}[c]{@{}c@{}}Right-hand\\      Galaxy\\      Smartwatch\\      in Reality\\      (User sees \\      the virtual\\      watch in the\\      VR environment)\end{tabular}}} &                                                                                                                 &                                                                                                            & \multirow{2}{*}{\begin{tabular}[c]{@{}c@{}}-VR Rright-hand\\      Controller\\       "A" Button\\      -Simple Reaction \\      Time\end{tabular}} \\ \cline{1-2}
D - VR-WT                                                             & \begin{tabular}[c]{@{}l@{}}-Indoor\\      -VR Simulated\\      Roadway Work Zone\\      -Simulated Traffic\\      -Ambient Noise\end{tabular}                                                        & \multicolumn{1}{c|}{}                                                                                                                     &                                                                                                                    & \multicolumn{1}{c|}{}                                                                                                      &                                                                                                                                                & \multicolumn{1}{c|}{}                                                                                                                                                                                                       &                                                                                                                 &                                                                                                            &                                                                                                                                                  \\ \cline{1-2} \cline{9-10} 
E - Vision-based                                                          & \begin{tabular}[c]{@{}l@{}}-Indoor\\      -VR Simulated\\      Roadway Work Zone\\      -Simulated Traffic\\      -Worker Engaged \\      in Maintenance Activity\\      - Ambient Noise\end{tabular} & \multicolumn{1}{c|}{}                                                                                                                     &                                                                                                                    & \multicolumn{1}{c|}{}                                                                                                      &                                                                                                                                                & \multicolumn{1}{c|}{}                                                                                                                                                                                                       &                                                                                                                 & \begin{tabular}[c]{@{}c@{}}One \\      Combination\\      (HAV)\end{tabular}                               & \begin{tabular}[c]{@{}c@{}}-Vision-based\\      Pose Tracking\end{tabular}                                                                        \\ \hline
\end{tabular}}
%\end{sidewaystable}
\end{table*}

\begin{figure}[ht!]
    \centering
    \includegraphics[width=\linewidth]{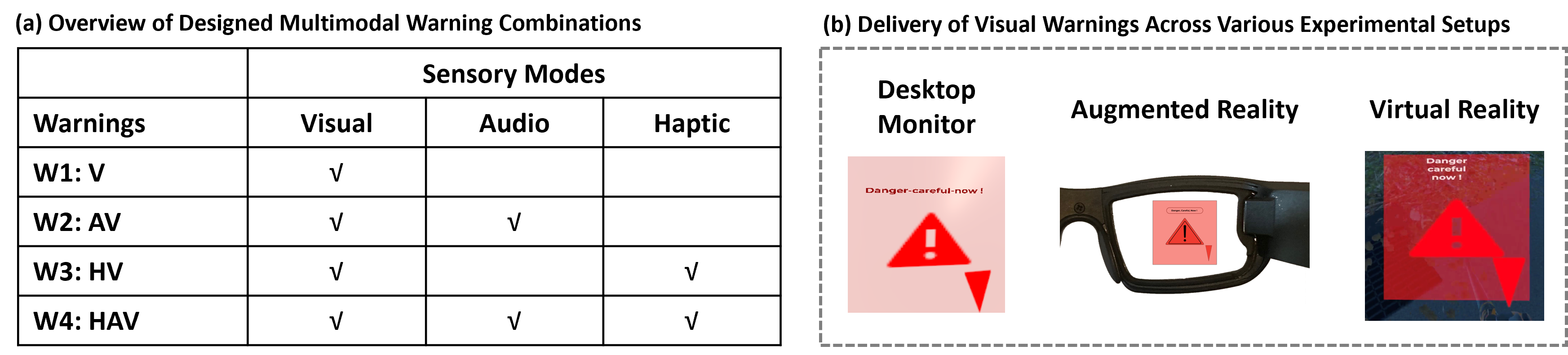}
    \caption{Overview of Designed Multimodal Combination of Warnings and Delivery of Visual Warning}
    \label{fig:notification}
\end{figure}

Figure \ref{fig:general} provides an overview of the hardware and software components of AR technology used in different experiments of this study. We used Vuzix Blade AR smart glasses and a Galaxy smartwatch, as shown in \ref{fig:general} (b), for this particular front-end design. We selected Vuzix Blade as they comply with ANSI Z87.1 standards and can be comfortably paired with prescription eyewear or shades. Featuring built-in audio modules and a display located on the right lens, these glasses ensured minimal obstruction of natural vision while being suited for industrial applications. The glasses, which operate on the Android operating system, facilitated the programming of the networking software and the design of our study-specific warning layout, shown in Figure \ref{fig:notification}(b). In terms of audio content, we used Bluetooth-connected earbuds in tandem with AR glasses to deliver audio cues directly to workers' ears. For haptic cues, the Tizen framework was utilized to develop the relevant software for the Galaxy smartwatch, enabling the delivery of haptic cues within the prototype's ecosystem. As a Samsung-specific platform, Tizen facilitated programming networking functionalities and other necessary algorithmic elements for seamless back-end and front-end communication in augmented reality glasses. 

\subsubsection{Virtual Reality Simulation}
The development of this simulation was centered on evaluating the suitability of virtual reality as a means of simulating AR warnings. For this purpose, the virtual reality simulation was designed to closely resemble the AR prototype developed. Similar layout design, visual cues, and audio frequency were implemented in the VR simulation to maintain consistency with the AR interface. We utilized the Oculus Quest 2 VR headset, as shown in Figure \ref{fig:general}(c) in the VR simulation. 
Furthermore, Unity game engine was utilized to develop pertinent software. We used the guidelines provided in the Manual on Uniform Traffic Control Devices (MUTCD) \cite{mutcd2006manual} to create a virtual work zone environment. This environment includes simulated traffic and highly detailed 3D models, closely resembling a real-world work zone. Figure \ref{fig:scene} illustrates some examples of the developed virtual environment. Furthermore, we employed the identical smartwatch component from the AR prototype in the VR simulation. During the simulation, users were able to simultaneously observe the virtual smartwatch on the screen while wearing the physical smartwatch on their wrist. This allowed us to simulate the exact interactions and user experience as those offered by AR technology. 

\subsubsection{Desktop-based Setup}
In addition to the AR and VR setups, we also developed a desktop-based replica of the AR warning to provide a baseline for comparison with the AR and VR interfaces. Using the software developed for the VR simulation in Unity, we made the necessary adjustments to create a desktop version of the AR warning mechanism. Figure \ref{fig:general}(a) illustrates the hardware and software setup for this desktop replica. The main difference between the desktop interface and the AR/VR settings is that visual and audio cues are delivered through the desktop display and speaker, rather than the AR glasses or VR headset. Algorithmic designs, developed software, and haptic feedback through the smartwatch remained exactly the same as in AR and VR technologies. 

\subsection{Study Design}
%In this section, we will outline the design and details of the experiments conducted. 
A total of five experiments were designed that fall into two distinct groups based on the reaction time measurement strategies. The first group, consisting of experiments A to D, aimed to quantify the reaction time to a number of warnings with different combinations of sensory modes in various settings and conditions using a simple reaction time approach. In contrast, during experiment E, we used a vision-based pose tracking metric to measure participants' reaction times. In this context, participants were exposed to scenarios that incorporated a task involving common work zone activities. Subsequently, we applied the non-intrusive vision-based methodology, as outlined in Section \ref{vision-based}, to calculate the reaction times. The task was carefully designed to simulate real-world scenarios and capture the naturalistic reactions of the participants. Further details of each set of experiments are provided in the following.

\subsection{Warnings}
Four different combinations of multimodal warnings were utilized, as shown in Figure \ref{fig:notification} (a). These comprise Visual (V), Audio Visual (AV), Haptic Visual (HV), and Haptic Audio Visual (HAV). Further information on each of these warnings, specifying the details of visual, audio, and haptic cues are provided in Table \ref{table:detail}. To ensure consistency in different settings, the layout of visual stimuli in all setups was designed to be identical, as illustrated in Figure \ref{fig:notification}(b). This uniform layout implementation aimed to minimize any potential variations or confounding factors introduced by differences in visual presentation across the different environments and experiments. Haptic stimuli were implemented using the Tizen Native framework, using a predefined pattern available in the API \cite{tizen}. Next, a high-pitched beep with a frequency of 44,100 Hz and a duration of 0.2 milliseconds was used as the auditory content of the warnings. It is important to note that all warnings were intentionally designed to trigger simultaneously, without any intentional delays, upon activation by the back-end system. This simultaneous triggering ensured that participants experienced multimodal warning stimuli in a synchronized manner, allowing a consistent evaluation of their reaction times in different multimodal warnings designed in this study.

\begin{figure}[ht!]
    \centering
    \includegraphics[width=\linewidth]{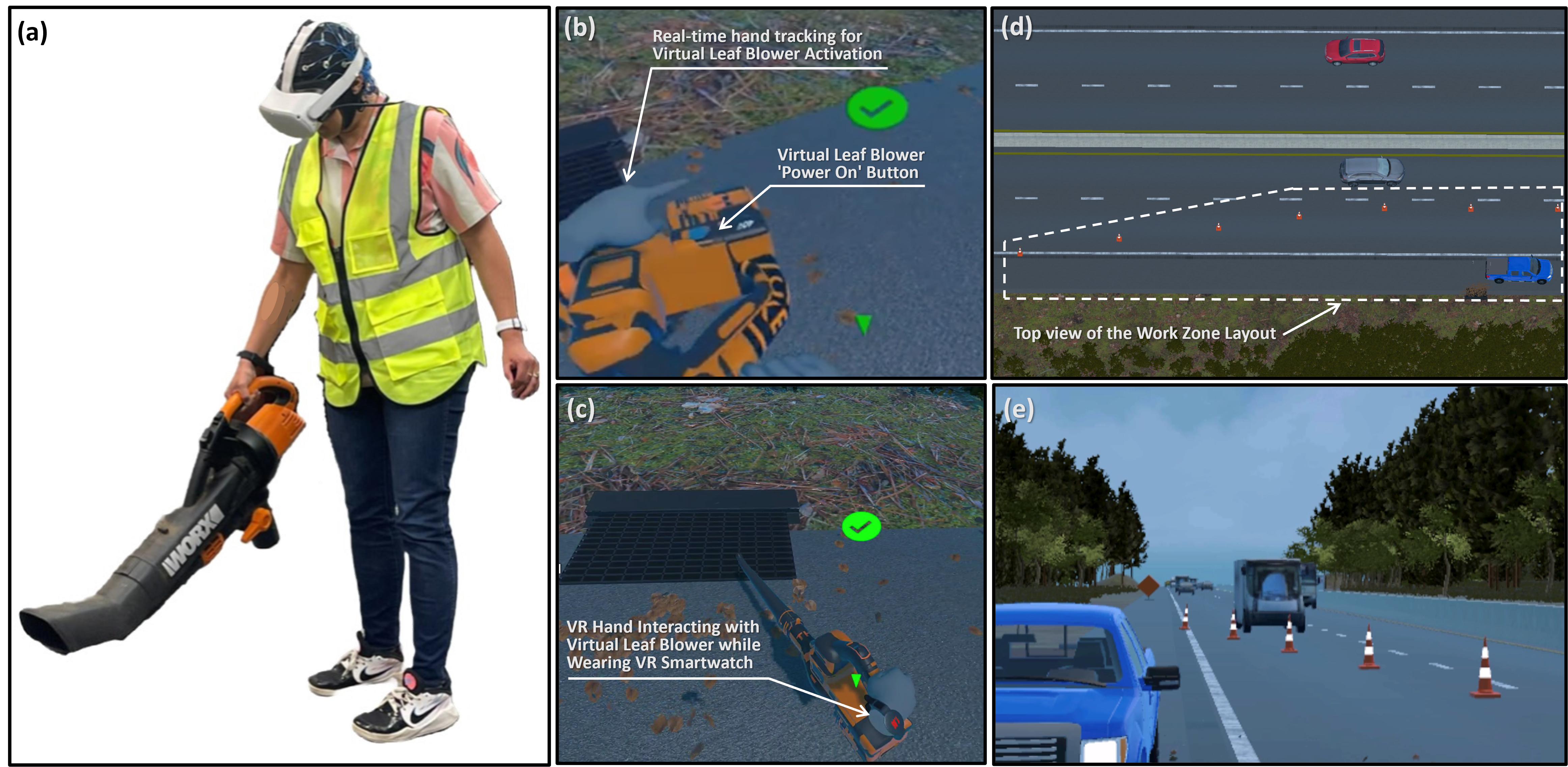}
    \caption{Virtual Reality Environment: (a) Lab setup showcasing the participant equipped with a VR headset, physically holding a real-life leaf blower, and wearing a smartwatch. (b) First-person VR perspective depicting the participant approaching the virtual leaf blower, preparing for activation. (c) Immersive and interactive VR environment captured from the participant's viewpoint, demonstrating the successful clearance of the drop inlet (d) Top-down view of the work zone layout, illustrating the operational area and placement of the virtual elements. (e) Near-miss scenario observed from the work zone perspective, showcasing dynamic traffic flow within the simulation}
    \label{fig:scene}
\end{figure}

\subsection{Experiment Procedure and Specifications}
The study protocol involved a carefully structured sequence of activities to ensure a comprehensive and ethical approach to our research. First and foremost, the ethical aspect of the study was prioritized by obtaining informed consent from participants. This step included transparently explaining the nature of the study, its goals, and the participants' roles. Participants were given the opportunity to ask questions and were fully informed about their rights and privacy protections. Following the consent process, we conducted an introductory session as part of our protocol. During this session, participants were greeted, and the purpose and objectives of the experiments were elaborated. We ensured that participants had a clear understanding of what to expect and what was expected of them throughout the study. In terms of logistics, we planned each experiment to be completed within a reasonable timeframe, with a maximum duration of up to half an hour per session. This study was conducted with the approval of the institutional review board (IRB No. 21-0357) of the University of North Carolina at Charlotte.

To maintain consistency across experiments A through D, a standardized Wizard of Oz (WOZ) scenario was devised and executed. In all these experiments, the WOZ scenario lasted for 45 seconds, during which warnings were presented to participants five times. The intervals between warnings were maintained consistently for each warning design throughout the experiments. However, to reduce the potential impact of a learning curve, these intervals varied for each type of warning. This setup ensured that the timing and frequency of warnings were systematically controlled and varied according to the specific sensory mode combinations being tested.

The scripted logic was used to implement the warning trigger points consistently across experiments A through D. The utilized intervals are as follows:

\begin{itemize}
    \item Visual warnings: triggered at 10, 20, 28, 33, and 36 seconds.
    \item Haptic Visual warnings: triggered at 15, 25, 28, 33, and 36 seconds.
    \item Audio Visual warnings: triggered at 17, 21, 28, 35, and 38 seconds.
    \item Haptic Audio Visual warnings: triggered at 12, 17, 22, 24, and 27 seconds.
\end{itemize}

Experiment B was conducted as the first experiment primarily due to logistical challenges. Given that this experiment took place outdoors, the main factors influencing this arrangement were weather-related constraints. To mitigate any potential influence from a learning curve, the sequence for conducting Experiments A, C, and D was deliberately randomized. Additionally, to address potential bias or confounding factors, the delivery and measurement order of reaction times for each type of warning in experiments A to D was also randomized. This randomization ensured a fair and unbiased assessment across the board. As an example, the reaction times of participant X were recorded in the sequence of AV, V, HAV, and HV, whereas the reaction times of participant Y were assessed in the order of V, AV, HV, and HAV. 

Moreover, Experiment E was carried out as the concluding experiment mainly due to logistical considerations, which included the necessity for special preparations, noticeable setup time and explanatory sessions. A WOZ-programmed scenario was carefully designed and executed consistently for all participants. The scenario had a duration of 1 minute and consisted of two iterations where the warning was triggered. The trigger points for the warning remained constant throughout the experiment. The warnings were triggered at 25 and 45 seconds. In particular, for Experiment E, the HAV warning was used because it encompassed all the stimuli under consideration. In the following, we provide further details and specifications of the experiments conducted.

\subsubsection{Experiment A: Establishing Reaction Time Baseline} 
This experiment was carried out using the developed desktop-based interface, as visualized in Figure \ref{fig:general}(a). The participants performed the experiment on a 21-inch display, sitting at a distance of 20 inches from the display, which was placed in front of a black background. A total of 32 participants (N = 32) completed the experiment, with an average age of 28.7 years (SD = 5.5) and an average of 3.4 years (SD = 0.9) of experience in the construction industry. Among the participants, 20 identified as male and 12 as female.

\subsubsection{Experiment B: Reaction Time to Augmented Reality Warnings in the Field}
This experiment was carried out in an outdoor setting in a temporary work zone that was specifically created for this study on the UNC Charlotte campus. The design of the work zone followed the guidelines outlined in the Manual of Uniform Traffic Control Devices (MUTCD) for short-duration work zones, as illustrated in Figure \ref{fig:general}(b). A total of 34 participants (N = 34) participated in this experiment, with an average age of 25.9 years (SD=5.1) and an average experience of 2.3 years (SD = 2.6) in the construction industry. Among the participants, 21 identified as male and 13 as female. The duration of the experiment for each participant ranged from 30 to 45 minutes. The experiment was carried out within the window of daylight hours, specifically from 11 am to 4 pm. In addition, weather conditions were assessed in advance using forecasting services to ensure the consistency of lighting conditions.

\subsubsection{Experiment C \& D: Reaction Time to AR Warnings in 
immersive VR Simulations} 
Both experiments were carried out using the simulated AR interface developed in virtual reality, as illustrated in Figure \ref{fig:general}(c). The VR environment was specifically designed to replicate a short-duration highway work zone based on the guidelines provided by the MUTCD. Figures \ref{fig:scene}(d) and (e) provide examples of the design of the virtual work zone used in the experiments. A total of 32 participants (N = 32) completed the study, with an average age of 28.7 years (SD = 5.5) and an average of 3.4 years (SD = 0.9) of experience in the construction industry. Among the participants, 20 identified as male and 12 as female. 

\subsubsection{Experiment E: Examination of Reaction Times to AR Warnings in a Mixed Reality Context Leveraging Pose Estimation} 
A total of 28 participants (N = 28) participated in the study, with an average age of 28.7 (SD = 5.6) and an average of 3.4 years of experience (SD = 0.9) in the construction industry. However, a total of 6 participants in the activity task encountered technical difficulties, simulation sickness, and other logistical problems during data collection and were unable to complete the experiment. As a result, the final count of participants in the experiment was 21 (N = 21), with an average age of 28.2 years (SD = 5.8) and an average industry experience of 3.1 years (SD = 1.1). 

In this experiment, our main goal was to replicate real-world scenarios commonly encountered in highway work zones. Our focus was specifically on the task of removing obstructions from drop inlets. To design our study, we considered the existing literature \cite{goenarjo2020cerebral,snyder2022aperture} that discusses the influence of physical activity intensity and cognitive load on reaction time. Based on this knowledge, we developed an obstruction removal task that required participants to engage in higher levels of physical exertion compared to other routine maintenance activities. Our intention was to simulate a task that is frequently encountered in the field of maintenance and operation of roadways. This task was chosen due to its practical significance and its common occurrence, allowing us to recreate real-world scenarios and evaluate the impact of warnings on participants' performance.

Our developed virtual work zone, depicted in Figures \ref{fig:scene}(d) and (e), served as the backdrop for this experiment. To enhance the realism and interactivity, we designed a mixed-reality interaction for  Experiment E. This approach allowed participants to engage with both physical objects in the real world and virtual objects within the virtual environment simultaneously. By combining elements from both realms, we aimed to create a unique and immersive experience for the participants. 

One key aspect of the mixed reality approach was the implementation of a leaf-blowing effect within the virtual environment. This effect was designed to simulate the action of using a leaf blower to clear leaves that obstruct a drop inlet. As participants entered the simulation, they were equipped with a physical leaf blower in their hands, mirroring the position and movements of the virtual leaf blower shown in Figures \ref{fig:scene}(a), (b) and (c). 

The task began with participants activating the leaf blower and directing it towards the obstructed drop inlet within the virtual environment. As they did so, the virtual reality environment featured a carefully designed blowing effect that effectively cleared the leaves positioned on top of the drop inlet. This dynamic and interactive task continued until all necessary warnings were delivered, and the administrator signaled the completion of the task.
\subsection{Reaction Time Measurement}
Experiments A through D focused on capturing the reaction time of the participants using an SRT strategy under various conditions. Each experiment consisted of five iterations of an SRT task in which participants were required to respond by pressing keys or buttons, as illustrated in Figures \ref{fig:general}(a), (b) and (c). However, the objective of Experiment E was to investigate whether multimodal AR warning, when presented, triggers any observable physical response that can be captured by pose tracking, as shown in Figure \ref{fig:general}(d). Our goal was to go beyond the traditional SRT approach and develop tasks that closely mimic real-world scenarios. In doing so, our objective was to assess the influence of real-time warnings on body movement and analyze that its connection with reaction time. In the following sections, we present a comprehensive description of the utilized strategies.
\subsubsection{Simple Reaction Time}
We used two similar approaches to measure reaction time in Experiments A-D. The methodology consisted of recording the time interval between receiving the warning triggering command from the server and pressing the designated capturing button. For the desktop interface, the designated button was the space button, while for AR and VR experiments it was the A button of the right controller of the VR headset. In the context of the AR experiment in the field, we improved the IoT network infrastructure and created custom software to facilitate the integration of virtual reality controllers for the purpose of capturing reaction times. To achieve this, we integrated the VR headset as an additional end point within the existing network. This configuration allowed us to utilize the controller in tandem with the AR prototype, facilitating the capture of reaction times. Communication latency between these endpoints was estimated to be less than 10 milliseconds \cite{sabeti2022toward}, which is considerably shorter than the average reaction time observed in humans \cite{jain2015comparative}. Therefore, the impact of this latency on the overall accuracy of the measurement was considered negligible.

\begin{figure}[t!]
    \centering
    \includegraphics[width=0.83\linewidth]{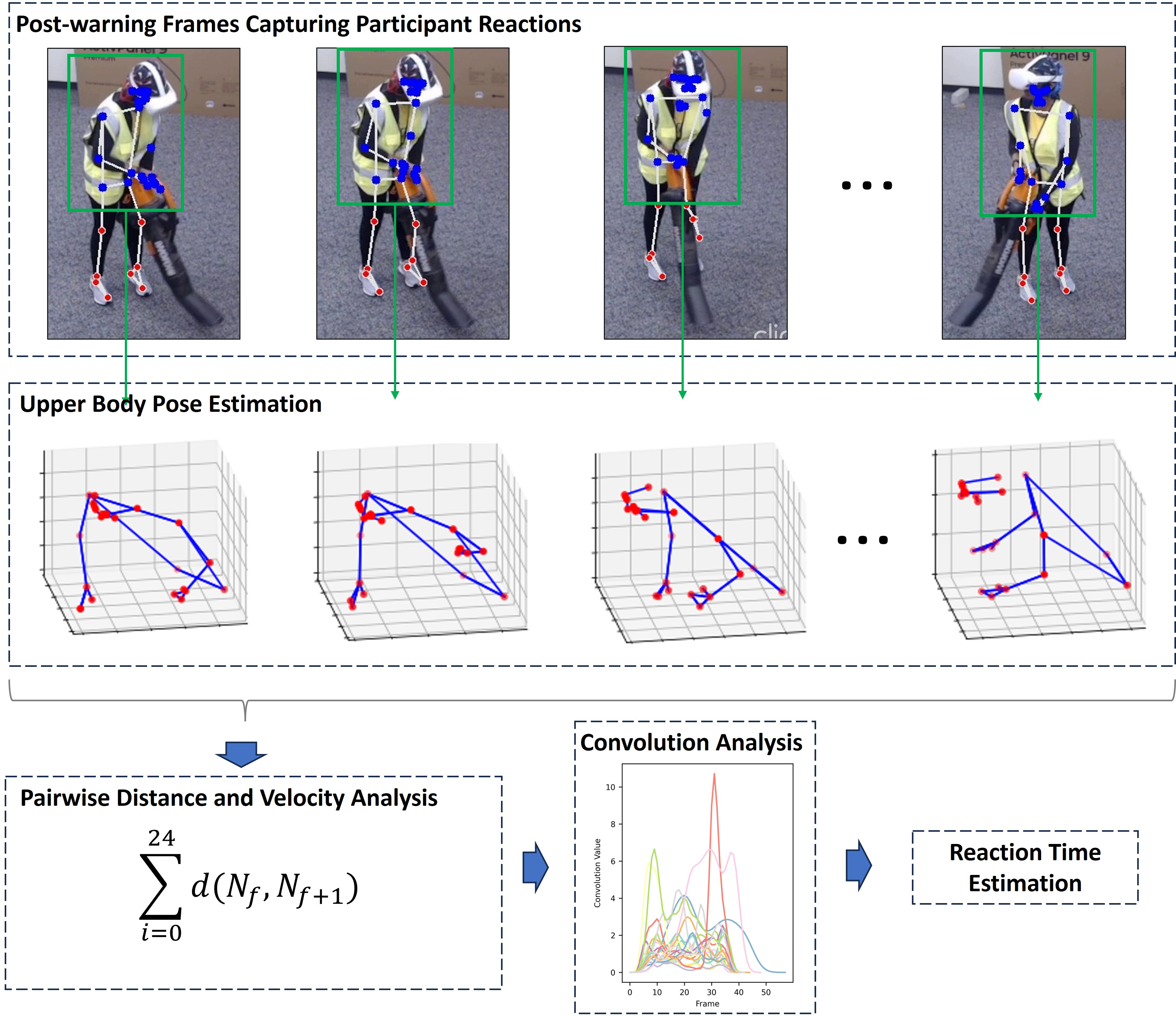}
    \caption{Proposed Vision-based Pose Estimation Methodology for Reaction Time Measurement}
    \label{fig:pose}
\end{figure}

\subsubsection{Vision-based Reaction Time}
\label{vision-based}
We utilized a Logitech camera, capable of capturing data at the rate of 30 frames per second (fps), to record the experiment and the reaction of the participants to warnings. To analyze body pose, we used the pose tracking API of ML Kit, an open source tool provided by Google \cite{googlemlkit}. This API offers a lightweight and flexible solution for the real-time detection and tracking of body poses from video streams and images. It provides a comprehensive 33-point skeletal map of the entire body, including facial landmarks, hands, and feet. In our analysis, we focused on the upper body landmarks, as we hypothesized that significant movement would occur primarily in this region. This hypothesis was derived from our observations and the relevant literature \cite{legg2021does}. Therefore, out of the 33 landmarks defined and tracked by the utilized pose estimation tracking algorithm, we limited our scope to the landmarks 0 to 24 according to \cite{googlemlkit}. Figure \ref{fig:pose} illustrates the proposed methodology for extracting reaction time from pose tracking.

To quantify the upper body movement, we extracted the raw coordinates of the landmarks from the output of the model applied to the video recordings. These landmarks were then filtered to select the considered joints for analysis. Using the coordinates of the selected landmarks, we calculated the total pairwise displacement between consecutive frames. This involved measuring the distance in the image space between each landmark and its previous 3D coordinate. By summing all these distances, we obtained the delta displacement, indicating the overall movement of the upper body. To find the velocity of the upper body movement, we divided the delta displacement by the corresponding frame duration. This calculation yielded the velocity of movement in terms of pixel distance per frame, allowing us to quantify the rate of movement. 

After calculating the upper body movement velocity, it was necessary to establish the reaction time pattern within the time series for further analysis. We defined the reaction time as the duration between the delivery of the warning and the onset of the reaction pattern exhibited by the participants. We used a Gaussian kernel to represent the reaction pattern within the velocity time series of subjects. This specific kernel has been used in the literature \cite{godfrey2008direct,uddin2020body} as an indicator of rapid reactions in the human body. Equation \ref{eq:pulse} illustrates the Gaussian kernel that is specifically used in our research. In this equation, $K(t)$ represents the Gaussian kernel at time $t$. The parameters $\mu$ and $\sigma$ control the shape and width of the Gaussian curve. $\mu$ represents the mean or center of the kernel, indicating the time at which the reaction pattern is expected to peak, while $\sigma$ represents the standard deviation, which determines the spread or width of the Gaussian curve. 

\begin{equation}
K(t) = amplitude \cdot \exp\left(-\frac{{(t - \mu)^2}}{{2 \sigma^2}}\right)
\label{eq:pulse}
\end{equation}

To analyze the collected time-series data and identify reaction-time patterns in the recorded velocity of body movements, we employed two pattern recognition techniques: convolution and wavelet analysis. Convolution plays a vital role in pattern recognition tasks within signal processing. It is commonly utilized to detect and extract patterns or features from signals by convolving a signal with a predefined pattern or filter \cite{begg2005machine}. The convolution operation highlights regions in the velocity signal where a pattern similar to the Gaussian kernel is observed, indicating the presence of the reaction pattern. Analyzing the resulting convolution output allows us to extract relevant features and information pertaining to the reaction time. Mathematically, the convolution of the time series with the kernel can be expressed as Equation \ref{eq:convoluion}:
\begin{equation}
y(t) = \int_{-\infty}^{\infty} x(t - \tau) k(\tau) \, d\tau
\label{eq:convoluion}
\end{equation}

To account for individual differences and ensure a personalized analysis, our study takes a within-subject approach. This involves determining the duration of each kernel based on the participant's prior performance in the HAV warning recorded in virtual reality conditions with the traffic scenario (VR-WT). By setting the kernel duration as the recorded reaction time, we aimed to accurately capture the temporal dynamics of each participant's response, individually. Through this customization, our aim is to capture the subtle nuances of each participant's reaction pattern. Additionally, to account for the anticipated agility in participants' reactions following the delivery of the warning, we further refined the kernel by setting its width to be one eighth of the total duration. This choice results in a steeper pulse shape, allowing us to capture the anticipated rapid changes in participants' response immediately after receiving the warning. The amplitude of the Gaussian kernel was standardized to a value of 1. Finally, to determine the time that corresponds to the maximum convolution value, we evaluated the function $y(t)$ over the desired time range and identified the time $t_{{max}}$ that satisfies the condition in \ref{eq:condition}. Using this approach, we can accurately identify the exact time at which the reaction pattern reaches its maximum intensity. The reaction time was then calculated as the duration between the delivery of the warning and the onset of the maximum convolution pattern observed, as shown in Equation \ref{eq:formula}.

\begin{equation}
\small t_{{max}} = \arg\max_t y(t) 
\label{eq:condition}
\end{equation}

\begin{equation}
\text{Reaction~Time} = t_{\text{max}} - \frac{t_{\text{HAV}}}{2} \label{eq:formula}
\end{equation}

In addition to convolution analysis, we also used wavelet analysis to further investigate patterns in the collected data. Previous studies \cite{uddin2020body} have suggested the application of wavelet analysis, specifically the Gaussian wavelet, for the analysis of body movements. Wavelet analysis decomposes a signal into simpler components using an algorithm similar to Fourier analysis. However, wavelet analysis is particularly effective in capturing transient behavior and discontinuities commonly observed in human movement signals. It enables a more accurate characterization of anomalies, pulses, and other transient events within the signal \cite{sato2006wavelet}. The equation for wavelet analysis is given by Equation \ref{eq:wavelet}. In this equation, $W(a,b)$ represents the wavelet transform of the signal $f(t)$ at scale $a$ and translation $b$. The symbol $\psi$ denotes the complex conjugate of the mother wavelet function $\psi$, and $\psi^(t)$ represents the complex conjugate of the scaled and translated wavelet function $\psi(t)$. The integral is computed over the entire real line from $-\infty$ to $\infty$. The scaling factor $1/a$ ensures the appropriate normalization of the wavelet transform. In our study, we utilized 2\textsuperscript{nd}-order Gaussian wavelets, similar to our kernel analysis, with the aim of uncovering any underlying patterns hidden within the data.
%\begin{equation}

%\scriptsize
%W(a,b) &= \int_{-\infty}^{\infty} f(t) \Psi^* \left(\frac{t-b}{a}\right) dt \\
%&= \frac{1}{\sqrt{a}} \int_{-\infty}^{\infty} f(t) \psi^*\left(\frac{t-b}{a}\right) dt

%\label{eq:wavelet}
%\end{equation}

{\tiny
\begin{equation}
\begin{aligned}
W(a,b) = \int_{-\infty}^{\infty} f(t) \Psi^* \left(\frac{t-b}{a}\right) dt \quad
= \frac{1}{\sqrt{a}} \int_{-\infty}^{\infty} f(t) \psi^*\left(\frac{t-b}{a}\right) dt 
\end{aligned}
\label{eq:wavelet}
\end{equation}
}

\section{Results}
\begin{figure}[b!]
    \centering
    \includegraphics[width=\linewidth]{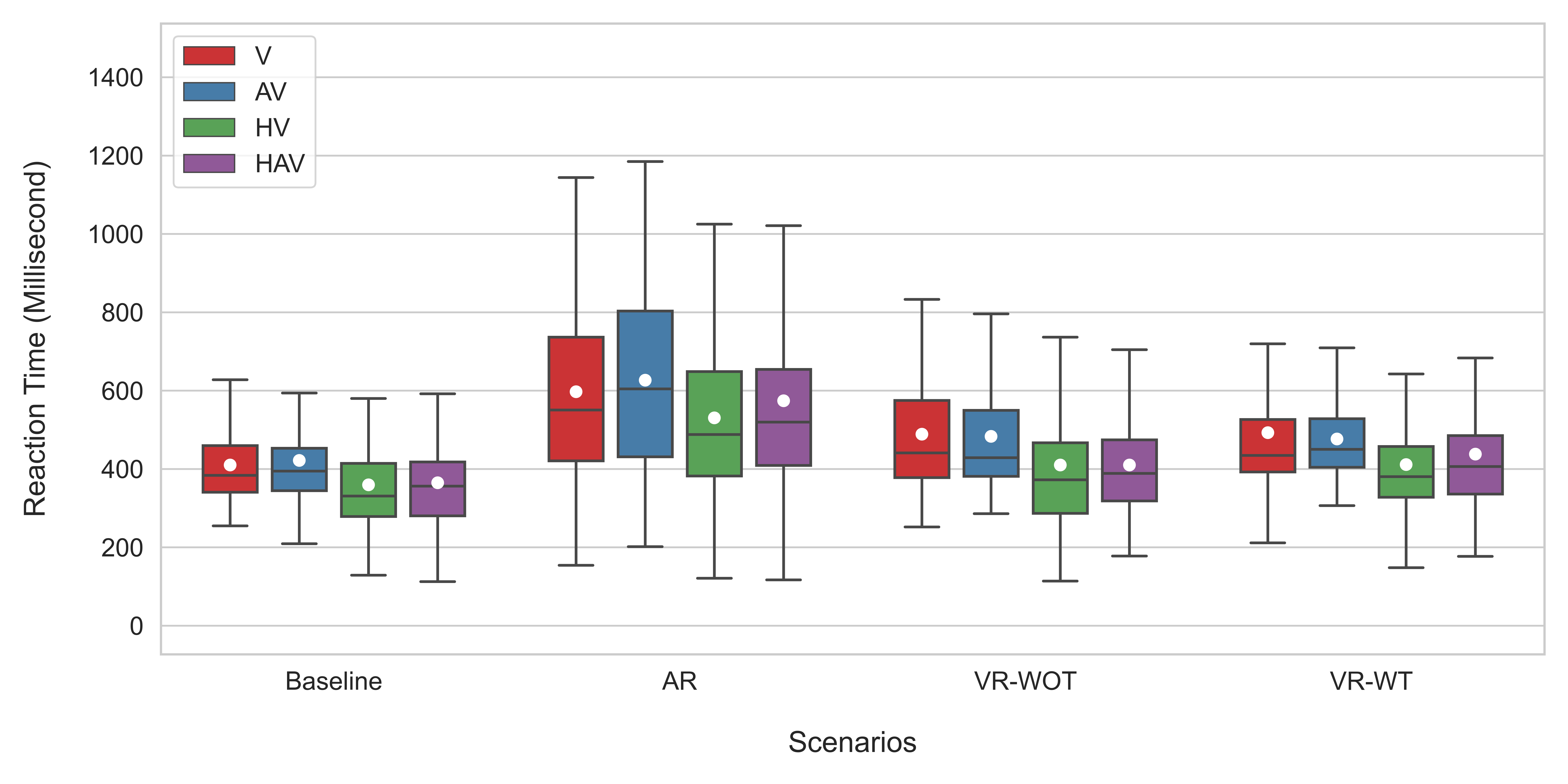}
    \caption{Reaction Times to Different Warnings (V: Visual, AV: Audio Visual, HV: Haptic Visual, HAV: Haptic Audio Visual) Across Different Settings (AR: Augmented Reality, WT: With Traffic, WOT: Without Traffic)}
    \label{fig:Comparison_Mode}
\end{figure}

\begin{table}[t!]
\centering

\caption{Reaction Times for Different Warnings Across Different Settings (AR: Augmented Reality, VR-WOT: Virtual Reality Without Traffic, VR-WT: Virtual Reality With Traffic)}
\label{Table:summary_Comparsion_Devices}

\renewcommand{\arraystretch}{1.18}
\resizebox{0.49\textwidth}{!}{
\begin{tabular}{c c c c c c}
\hline
& & \multicolumn{1}{c}{Baseline} & \multicolumn{1}{c}{AR} & \multicolumn{1}{c}{VR-WOT} & \multicolumn{1}{c}{VR-WT} \\
\hline
\multirow{2}{*}{V} & Average (ms) & 410 & 597 & 489 & 493  \\
& SD & 105 & 232 & 159 & 177 \\
\hline
\multirow{2}{*}{AV} & Average (ms) & 422 & 627 & 483 & 477 \\
& SD & 122 & 249 & 162 & 108 \\
\hline
\multirow{2}{*}{HV}& Average (ms) & 359 & 530 & 410 & 411  \\
& SD & 145 & 245 & 184 & 149  \\
\hline
\multirow{2}{*}{HAV}& Average (ms) & 365 & 574 & 411 & 438  \\
& SD & 149 & 273 & 127 & 154  \\
\hline
\end{tabular}}
\end{table}

We summarize the results in two sections. The first details participants' reaction times to different multimodal warnings across the experiments. The following section includes the results of investigating the physical responses exhibited by participants following the delivery of the warning using our vision-based pose tracking approach.

\subsection{Impact of Warning Modalities on Reaction Time}

Figure \ref{fig:Comparison_Mode} provides a comprehensive summary of the results obtained from experiments A to D. This graph offers insight into the reaction times associated with different multimodal warnings in various experiments and scenarios. Several notable patterns and trends emerge from this graph. It can be observed that, on average, the reaction times to AR warnings in the real-world setting (Experiment B) are longer and exhibit higher variability compared to the baseline (Experiment A), which have shorter duration and lower variability. Table \ref{Table:summary_Comparsion_Devices} further supports this observation, indicating that the reaction times to AR-delivered warnings in real-world setting (Experiment B) are consistently higher with greater variability compared to both baseline values and simulated warnings. These discrepancies in reaction times can be attributed to the inherent differences between real-world testbeds and the controlled indoor environment where the baseline experiment was conducted. The real-world setting introduces numerous cognitive distractions that are absent in indoors, potentially impacting participants' response times and leading to the observed variability. The results of the statistical comparison through t-tests summarized in Table \ref{tab:stasistical_summary}(a) further support this observation, confirming the statistical differences between the means of the collected samples.

%\lipsum[1] % Some dummy text before the table
\begin{table}[hb!]
\centering
\renewcommand{\arraystretch}{1.18}
\caption{Comparison of Reaction Times Across Experimental Setups and Different Warnings: t-test Analysis and Obtained p-values}
\resizebox{0.49\textwidth}{!}{
%\scalebox{0.9}{

\begin{tabular}{ccccclccccc}
                     &        &          &      &        &  &                           &     &      &      &      \\ \hline
\multicolumn{5}{c}{ (a) Experimental Setup}                   &  & \multicolumn{5}{c}{(b) Warning Type}                   \\ \cline{1-5} \cline{7-11} 
                     &        & Baseline & AR   & VR-WOT &  &                           &     & V    & AV   & HV   \\ \cline{1-5} \cline{7-11} 
\multirow{3}{*}{V}   & AR     & 0.000     &      &        &  & \multirow{3}{*}{Baseline} & AV  & 0.278 &      &      \\
                     & VR-WOT & 0.000     & 0.000 &        &  &                           & HV  & 0.015 & 0.001 &      \\
                     & VR-WT  & 0.000     & 0.000 & 0.839   &  &                           & HAV & 0.422 & 0.075 & 0.138 \\ \cline{1-5} \cline{7-11} 
\multirow{3}{*}{AV}  & AR     & 0.000     &      &        &  & \multirow{3}{*}{AR}       & AV  & 0.356 &      &      \\
                     & VR-WOT & 0.000     & 0.000 &        &  &                           & HV  & 0.000 & 0.000 &      \\
                     & VR-WT  & 0.000     & 0.000 & 0.667   &  &                           & HAV & 0.002 & 0.000 & 0.744 \\ \cline{1-5} \cline{7-11} 
\multirow{3}{*}{HV}  & AR     & 0.000     &      &        &  & \multirow{3}{*}{VR-WOT}   & AV  & 0.335 &      &      \\
                     & VR-WOT & 0.007     & 0.002 &        &  &                           & HV  & 0.000 & 0.000 &      \\
                     & VR-WT  & 0.000     & 0.007 & 0.958   &  &                           & HAV & 0.004 & 0.011 & 0.117 \\ \cline{1-5} \cline{7-11} 
\multirow{3}{*}{HAV} & AR     & 0.000     &      &        &  & \multirow{3}{*}{VR-WT}    & AV  & 0.764 &      &      \\
                     & VR-WOT & 0.000     & 0.000 &        &  &                           & HV  & 0.000 & 0.000 &      \\
                     & VR-WT  & 0.000    & 0.003 & 0.082   &  &                           & HAV & 0.000 & 0.000 & 0.997    \\ \hline
\end{tabular}
}
\label{tab:stasistical_summary}
\end{table}
%\lipsum[2] % Some dummy text after the table

Another intriguing observation from the findings pertains to the comparison of simulated AR warnings in virtual reality with and without the presence of traffic (Experiments C and D). The outcomes illustrated in Figure \ref{fig:Comparison_Mode} demonstrate that the reaction times to these simulated warnings exhibited similar patterns, regardless of the presence of traffic and the simulated ambient noise. The average recorded reaction time values in both cases fell between the reaction times observed in the real-world AR and the controlled baseline environment. This observation is further supported by the results of a t-test provided in Table \ref{tab:stasistical_summary}(a), indicating that the reaction times were not statistically different. This finding holds true across all types of utilized warnings, even for those that did not include audio stimulus, highlighting that the inclusion of ambient noise and traffic behavior in the experiment did not noticeably impact participants' reaction times when interacting with AR warnings in the simulated environment. 

Furthermore, the results of the experiments, as shown in Table \ref{tab:stasistical_summary}(a), indicate that there are significant statistical differences between the mean reaction time values collected in the real-world and simulated environments. The t-test conducted between the different versions of simulated AR warnings (with and without traffic) demonstrates that the reaction times observed in the VR simulations are statistically different from the reaction times recorded in the real-world environment. This indicates that virtual reality, as utilized in this study, was not able to accurately replicate the intricacies and complexities of the outdoor environment when measuring reaction times in AR applications. The reaction times collected in the virtual reality simulations, relative to the baseline measurements, are closer to the real-world measurements of reaction time to AR. However, it is important to acknowledge that they still show deviations from the actual values observed in the real-world setting. We will discuss this observation in detail in the Discussion section. 
\begin{figure}[t!]
\centering
\includegraphics[width=\linewidth]{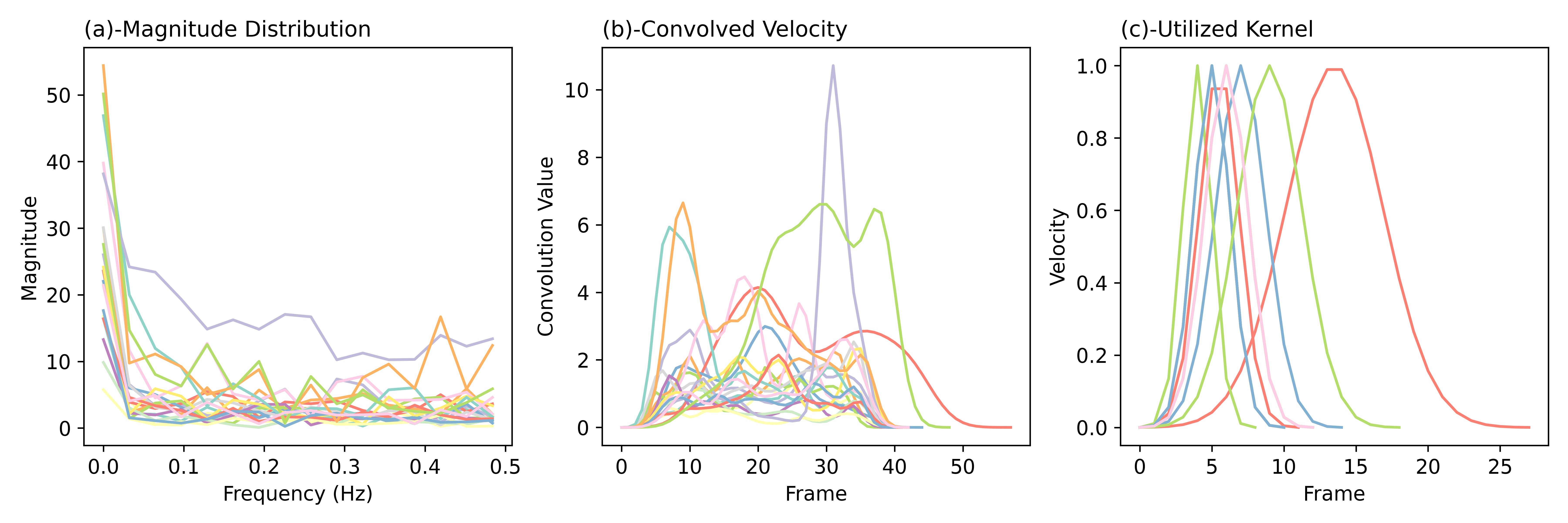}
\caption{Specifications of the Adopted Time-series Analysis in the First warning: (a) Velocity Time-series of Upper Body Cumulative Joint Movement in Participants, (b) Magnitude Distribution Per Frequency and (c) Utilized Individual Kernels for Each Participant Based on the Recorded Baselines}
\label{fig:Vision_based_heatmap_heavy1}
\end{figure}

Comparing the impact of different multimodal warnings on RTs provides another avenue for analyzing the collected data. As illustrated in Figure \ref{fig:Comparison_Mode}, the results indicate that, on average, warnings with a combination of visual and haptic cues (HV) triggered faster reaction times across different setups in Experiments A to D, as shown in Table \ref{Table:summary_Comparsion_Devices}.  Furthermore, summarized results in Table \ref{tab:stasistical_summary}(b) indicate that the difference in reaction times between HV and V, as well as HV and AV warnings, is statistically significant, indicating that HV warning consistently triggered faster reaction times compared to other warnings with only visual or audio components.However, as results in Figure \ref{fig:Comparison_Mode} and Table \ref{tab:stasistical_summary}(b) indicate, the difference between HV and HAV warnings across all experiments was statistically insignificant. This implies that the HAV warning triggered statistically similar reaction times to those of the HV warning. This outcome contradicts the general expectation that the HAV design would outperform other designs. We will discuss this finding in the Discussion section in greater detail.

\begin{figure}[t!]
\centering
\includegraphics[width=\linewidth]{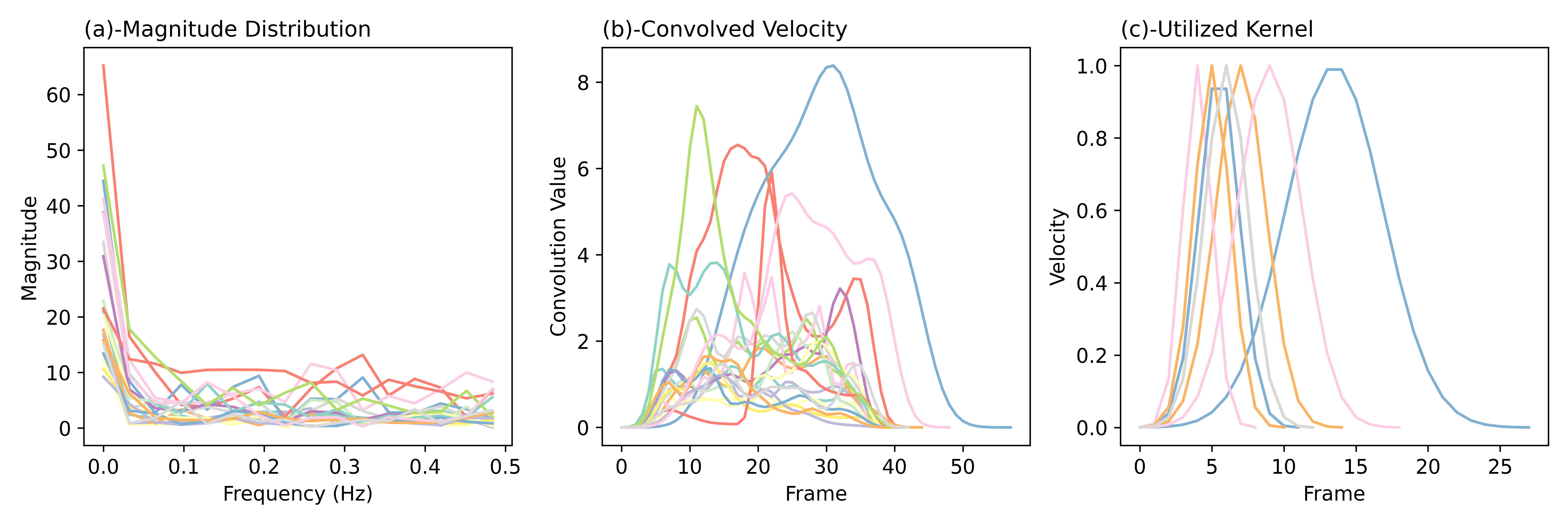}
\caption{Specifications of the Adopted Time-series Analysis in the Second warning: (a) Velocity Time-series of Upper Body Cumulative Joint Movement in Participants, (b) Magnitude Distribution Per Frequency and (c) Utilized Individual Kernels for Each Participant Based on the Recorded Baselines}
\label{fig:Vision_based_heatmap_heavy2}
\end{figure}

\subsection{Vision-based Reaction Time Measurement}
In this section, we present the results of the proposed vision-based approach to quantify the reaction time. We set the duration of the Gaussian kernel as the reaction time of each participant to the HAV warning recorded under the simulated AR condition in VR with traffic (VR-WT) in the previous steps. To focus our investigation, we restricted our search to a specific timeframe of 30 frames or 1 second following the delivery of the warning in the video recordings. This time limit was chosen based on the observation that 99.7\% of the reaction times collected for the HAV warning in the VR-WT scenario were within the range of 438 milliseconds plus three times the standard deviation (154 milliseconds as shown in Table \ref{Table:summary_Comparsion_Devices}). Therefore, we deemed this timeframe to be a suitable data-driven representative considering previous observations.

\begin{figure}[ht!]
\centering
\includegraphics[width=\linewidth]{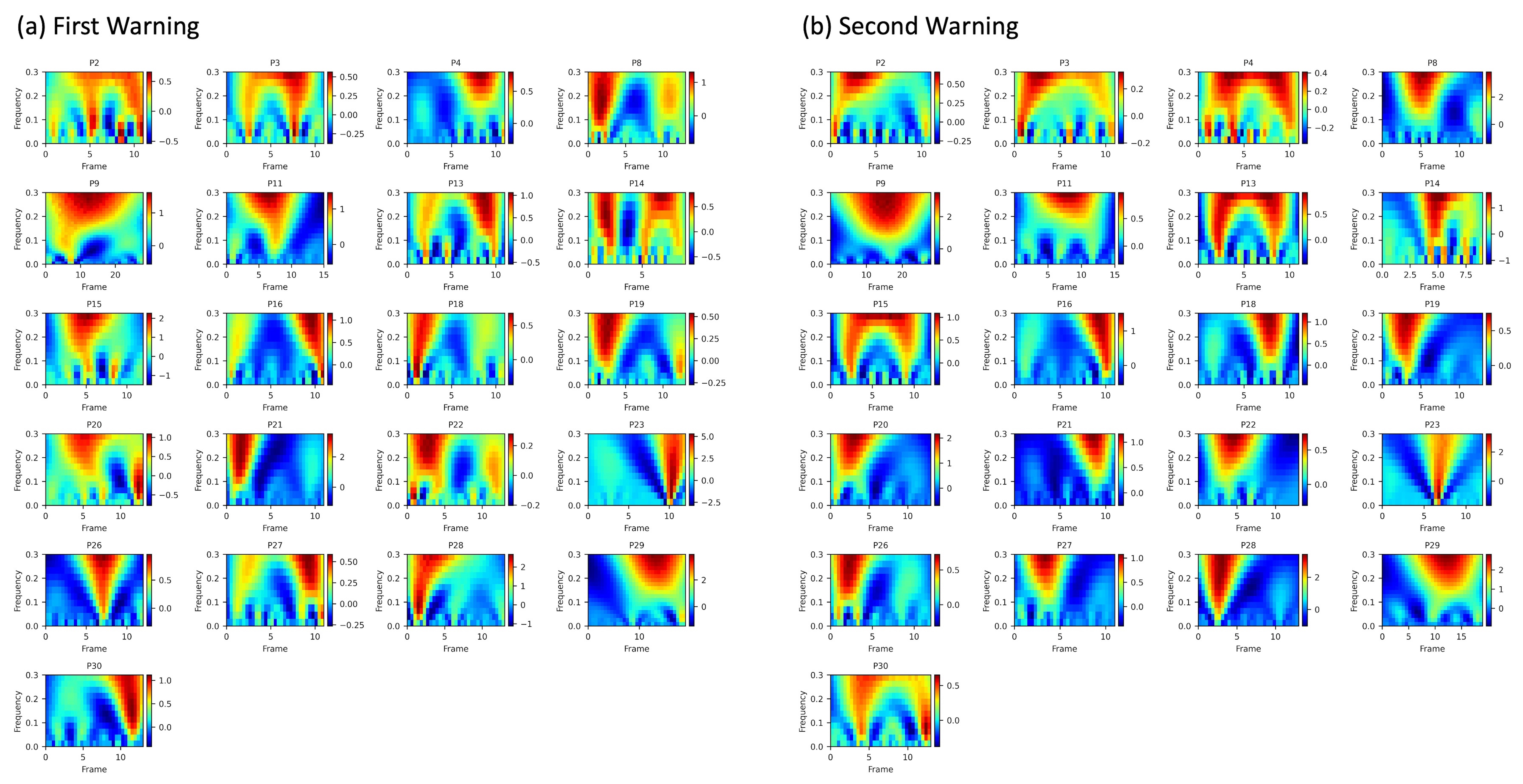}
\caption{Wavelet Analysis Results on Velocity Time-series of Cumulative Upper Body Movement for Each Participant in Response to the (a) First and (b) Second Warning}
\label{fig:Vision_based_heatmap_heavy1_wavelet}
\end{figure}

We then applied our convolution analysis approach to the filtered time series and summarized the results in Figure \ref{fig:Vision_based_heatmap_heavy1}. This figure demonstrates the outcomes of the proposed methodology applied to responses to warnings 1 and 2, respectively. To calculate the strength of the frequency content of the velocity time series, we used Fast Fourier Transform (FFT), decomposed the signal into frequency domain, and calculated the magnitude of each frequency strength. The results of this signal processing approach are summarized in Figures \ref{fig:Vision_based_heatmap_heavy1}(b) and \ref{fig:Vision_based_heatmap_heavy2}(b). Figures \ref{fig:Vision_based_heatmap_heavy1}(c) and \ref{fig:Vision_based_heatmap_heavy2}(c) also illustrate the individualized kernels used in the analysis. These figures highlight a consistent trend in the strength distribution across different frequencies in the post-warning velocity of upper body movement. Additionally, the results indicate a similar one-peak convolution distribution among different participants. 

Furthermore, we performed wavelet analysis on the calculated reaction times to the first and second warnings in Experiment E and present the results in Figures \ref{fig:Vision_based_heatmap_heavy1_wavelet} (a) and (b), respectively. These figures illustrate the coefficients of the frequencies of the processed signals, revealing a similar one-peak distribution in almost all participants with very few exceptions in both post-warning upper body movement velocities. These findings demonstrate the efficacy of our vision-based approach in capturing and analyzing reaction times patterns in different participants. The consistent trend observed in the energy distribution and coefficients across different frequencies and participants supports the validity of the proposed method.

\begin{table}[hb!]
\centering
\renewcommand{\arraystretch}{1.18}
\caption{Summary of The Collected Reaction Times In milliseconds Using Vision-based Metric}
\label{table:comparison_vision_based}
\scalebox{0.99}{

\begin{tabular}{c c c}
\hline
  & \multicolumn{1}{c}{First Warning} & \multicolumn{1}{c}{Second Warning} \\
\hline
Average (ms)               & 490   & 370   \\
SD                         & 330   & 220   \\
 Paired t-test with HAV in VR-WT & 0.36   & 0.42   \\
Paired t-test with Each Other & \multicolumn{2}{c}{0.19} \\
\hline
\end{tabular}
}
\end{table}

\begin{figure}[t!]
\centering
\includegraphics[width=\linewidth]{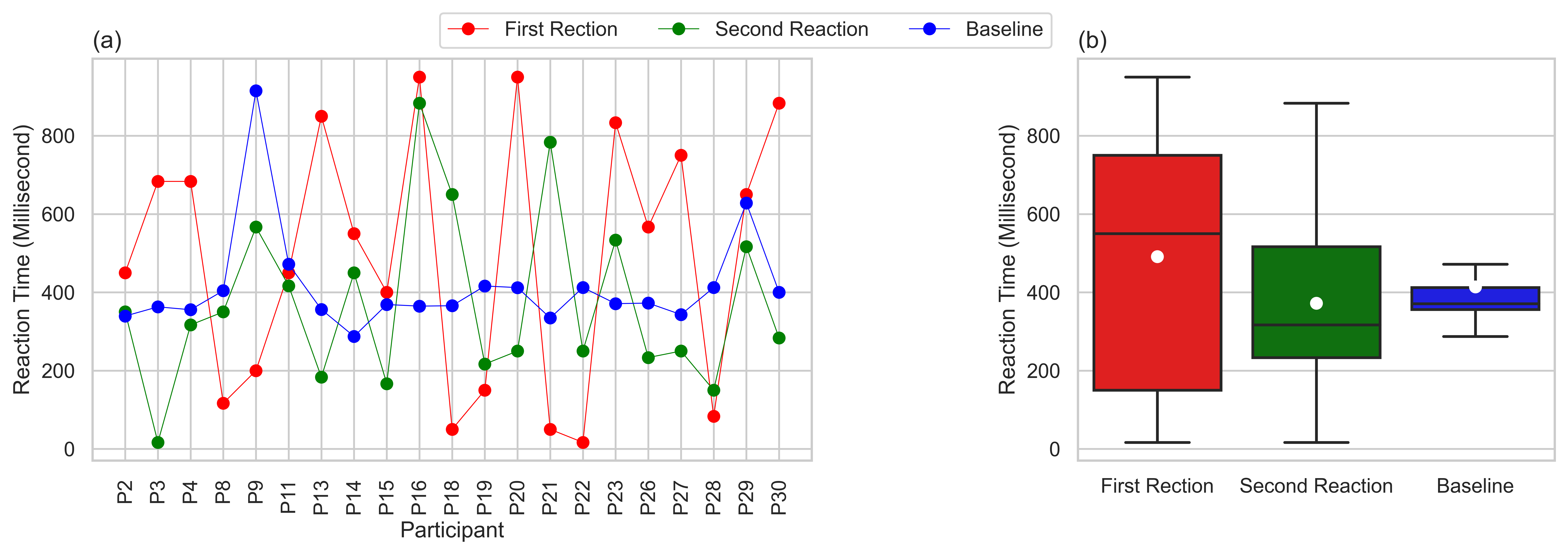}
\caption{(a) Calculated Reaction Times for Each Participant using the Vision-Based Metric in the First and Second warning along with the Recorded Baseline and (b) Box plot of The Recorded Values}
\label{fig:comparison_vision_based_point}
\end{figure}

Finally, Table \ref{table:comparison_vision_based} provides a summary of the results obtained from our vision-based approach to quantify the reaction time. This table presents the average and standard deviation of the reaction times calculated for each participant in the first and second warnings. To validate the effectiveness of our strategy, we used paired t-test analysis to statistically compare reaction time values measured with the vision-based approach and RTs to HAV warnings in the VR-simulated experiment (Experiment D).
The results of the t-tests indicate that the vision-based reaction times are not statistically different from the baseline, suggesting that our vision-based approach yields results comparable to the SRT metrics. Furthermore, we performed paired t-tests between the reaction time measurements in the first and second rounds of vision-based metrics. The results of these t-tests revealed that our strategy resulted in reaction times in both rounds that are statistically comparable. These findings are visualized in Figure \ref{fig:comparison_vision_based_point}, where the reaction times collected in the first and second rounds of the vision-based strategy, as well as the baseline, are presented. It is worth noting that although the standard deviation of the vision-based metric was higher than the baseline, the average reaction times were statistically comparable. The consistency of results between the vision-based and the SRT metrics, along with the stability of vision-based measurements across multiple iterations, further highlights the potential of computer vision and pose estimation techniques to investigate real-time reaction times. 

\section{Discussion}

\subsection{Effects of Combinations Warning Modalities}
Through the review of previous studies, we consistently found that the reaction time benefits of using multimodal warnings outweigh those of unimodal designs across various contexts, including transportation applications \cite{ jia2019effect,geitner2019comparison,lee2023investigating}. We have also observed that one recurring theme in reaction time studies is understanding how different multimodal warnings impact end-users in terms of response time, sensory overload, and user experience. In our specific use case, where AR is a fundamental component of the safety system, we examined a unimodal visual warning communicated through the AR glasses. As expected, this unimodal warning was outperformed by the various multimodal cues we examined. However, what caught our attention was the remarkable performance of the HV warning, which consistently outperformed all other combinations of multimodal warnings across different scenarios, even surpassing the HAV warning that incorporates three cues (i.e., haptic, audio, and visual).

One potential explanation for the poor performance of HAV warning relative to HV may be attributed to sensory overload. Several studies \cite{douce2020sensory, katic2015system,wueffect} have emphasized the impact of this phenomenon on human perception and response. Consequently, the use of more cues in the design of multimodal warnings does not necessarily lead to faster response times in end-users. A careful and context-aware evaluation is essential when making decisions about the best multimodal warning designs in specific contexts. Like this work, studies in other fields have also attempted to answer this question within their respective contexts. For example, Edet et al. \cite{edet2022evaluation} conducted a study in the context of autonomous vehicles in agriculture. In pursuit of their objective, their study employed a virtual simulation of an autonomous sprayer to measure the operator's reaction time to seven warning methods, which were presented either individually or in combinations of visual, auditory, and tactile sensory cues. They discovered that among seven warning methods, a combination of tactile and visual methods proved to be the most effective for their specific situation. Their findings also align with the conclusions drawn in our study regarding the performance of multimodal warnings.

Another plausible explanation for the relatively underwhelming performance of the multimodal warnings that include the audio sensory mode, such as HAV and AV could be attributed to the unique conditions of our study, where we employed a specific audio pattern with a particular frequency. Several previous studies \cite{williams2014effects,navarro2023mathematical} have already established that the characteristics of audio cues can influence the reaction time. It is important to note that our main focus in this study did not include investigating the impact of the specificity of audio or other cues (visual and haptic) on the reaction time of workers. However, we find it intriguing for future studies to focus on quantifying how different audio warning designs, as well as alternative designs and patterns of haptic and visual cues, could potentially impact the reaction time of workers.

\subsection{Implication of Reaction Time on Work Zone Design}
Navigating the dynamic work environment on highways presents substantial challenges to workers, particularly in responding to unexpected intrusions. This complexity arises primarily due to the limited time available for response to hazards, as highlighted in the literature \cite{fyhrie2016work,nnaji2020improving}. In the United States, the Manual on Uniform Traffic Control Devices (MUTCD) \cite{mutcd2006manual} has been the long-standing guide for designing work zones on highways. According to this manual, the design of geometrical characteristics of highway work zones must take into account various environmental and contextual factors, including posted traffic speeds, geometrical characteristics of the roadways, lighting conditions, etc.

The MUTCD manual suggests providing an optional buffer space within work zones. However, in practice, this buffer is often not provided. In such situations, on a highway with a posted maximum speed limit of 65 mph, workers should be positioned approximately 1,037 feet away from the first cone in the nose of the work zone. Consequently, in the event of a collision with the initial cone, a vehicle traveling at this speed would take approximately 10 seconds to reach the work area. This narrow time frame of 10 seconds underscores the urgency of workers' reactions to such intrusions. However, it is essential to acknowledge that not all intrusions originate from the front of the work zone; scenarios involving sudden swerving or lateral intrusions offer even less reaction time. These types of intrusions are often among the most hazardous and can result in fatal incidents within work zones \cite{mosunmola2022countermeasures}.

Given the limited time available for workers to respond to dangers, current technologies appear to be inadequate to address this challenge, particularly in scenarios involving lateral intrusions. Review studies \cite{nnaji2020improving, nnaji2020case} have indicated that while existing intrusion alert technologies have the potential to improve worker safety, they have shortcomings, such as insufficient reaction times, persistent false alarms, lengthy setup procedures, and inaudible alarms from distance. Furthermore, previous research \cite{nnaji2018developing,gambatese2017work} has shown that, through rigorous in-field testing, workers' reaction times to existing technologies can be as long as 2-3 seconds at a minimum. Furthermore, these studies have identified that the relative proximity of workers to alert technology and their orientation (e.g., whether they are in front or behind alert technologies) significantly influence their response times. In contrast, our findings suggest that personalized AR-based warnings substantially reduce reaction times, all while eliminating location-based constraints associated with one-size-fits-all technologies. Specifically, our results demonstrate that by utilizing multimodal AR-based warnings, workers can achieve an average reaction time of less than 0.6 seconds, providing them with a crucial advantage in effectively responding to rapidly evolving hazards.

\subsection{Strengths and Weaknesses of using VR for AR Simulation:}
As discussed in the literature review, VR has gained increasing recognition as an interactive and immersive simulation tool, particularly for replicating scenarios that pose logistical, financial, and experimental challenges. One of the primary promises of VR lies in its ability to accurately model physical, behavioral, and contextual environments of the real world that are pivotal for successful simulations, as shown in several studies \cite{hasan2021distributed,harris2021exploring,mills2020virtual}. However, while virtual reality has demonstrated success in various domains and applications, the need for benchmarking and data-driven comparisons with real-world testbeds becomes increasingly consequential as reliance on simulation grows, especially in critical areas such as safety. Although the reaction times captured in our VR simulation experiments did not achieve absolute validity — in that they did not statistically align with real-world measurements — they demonstrated relative validity by reflecting the underlying trends observed in real-world. In a comparison of the various warnings presented during this study, similar patterns of variability were evident in both VR-based settings and real-world environments. Specifically, AV warnings were identified as the least effective, while HV warnings were the most potent in prompting reactions from participants. This underscores the capability of VR to simulate highway work zones with a commendable level of fidelity, suggesting promising avenues for further research in an area that has seen limited attention.

Another important consideration when using VR simulations is the potential influence of audiospatial attention on reaction times. Previous studies have indicated that the way audio is delivered in immersive environments, such as through stereo versus binaural sound, can have significant implications for how users perceive the virtual environment \cite{hoeg2017binaural}. This aspect could be relevant to our experiments, as the reaction times collected in both VR-based scenarios (with and without traffic) surprisingly did not show a statistically significant difference. This means that even though simulated traffic streams and ambient noise were added to the VR-WT scenario, the impact of the added audio was not reflected in the collected reaction times. Therefore, it would be intriguing to further study how audiospatial attention affects reaction times to simulated AR-based warnings in future studies. This investigation could provide valuable information on optimizing audio cues in VR simulations for enhanced realism and effectiveness, particularly in safety-critical applications.

\subsection{Computer Vision for Reaction Time Measurement: Promise and Applications}
Non-intrusive safety measures are gaining popularity across various domains, particularly in the context of safety assessment \cite{fang2020computer}. These platforms have opened opportunities for a wide range of applications, including evaluating worker safety \cite{khan2022fall,guo2021computer}. However, to the best of our knowledge, no prior study has specifically focused on assessing the reaction times of highway workers using vision-based mechanisms. The proposed vision-based approach offers valuable information on the reaction abilities of high-risk workers by enabling the continuous assessment and evaluation of their cognitive and physiological responses through computer vision. It captures and analyzes relevant kinematic data in real-time, providing an understanding of how workers react and adapt to various situations. Importantly, this non-intrusive method minimizes operational disruptions and facilitates the continuous collection and assessment of data, allowing for more natural and realistic evaluations. Therefore, the proposed approach provides the capacity to identify potential risks, improve safety protocols, and improve worker performance. It could also be used to create customized training programs and precise interventions for workers who might demonstrate compromised reaction capabilities in high-risk environments and scenarios.

\subsection{Limitations and Future Work}
Although it was found that the auditory cue contribution may not be necessary in the warning design, this conclusion is based on the assumption of a constant frequency for the audio. Future studies could explore how variations in the design of the audio module, such as different frequencies or patterns, could affect reaction times and user responses. Additionally, future research can expand on the task design used in this study to other common activities or scenarios. This could involve investigating the relationship between cognitive load, physical participation, and reaction time in different contexts.

\section{Conclusion}
This paper evaluated the potential of augmented reality warnings in the context of roadway work zone safety by quantifying the reaction times to different multimodal warnings. Through five different experiments we measured the reaction times in real-world, indoor, and Virtual Reality simulated settings using Simple Reaction Time (SRT) and a vision-based approach. The haptic visual warning triggered the fastest response on average among the participants and produced measurements comparable to those of the audio haptic visual warning. Moreover, both of these warnings significantly outperformed visual and audio visual warning in terms of reaction time. Our findings also reveal that, on average, the reaction time to augmented reality warnings in real-world scenarios was longer, with greater variability compared to the baseline of desktop warnings and simulated AR in virtual reality. Also, VR-simulated warnings were not significantly shorter than AR warnings in real world. Furthermore, we observed a noticeable difference in reaction times between AR warnings and the baseline desktop version under different conditions. This emphasizes the importance of accounting for the real-world performance of workers when designing AR-oriented safety solutions. We also used a vision-based real-time pose tracking, to quantify reaction time. We used a within-subject approach to compare the individual-level reaction times with the baselines obtained from previous experiments. Our findings demonstrated the statistical comparability of the vision-based metric with the SRT-based metrics at an individual level.

The findings of this study provide future reference for both researchers and practitioners who are interested in leveraging AR, VR simulation, and reaction time in the context of safety. Furthermore, the proposed vision-based reaction time measurement approach has the potential for wider applications in assessing and monitoring the performance of workers whose reaction capabilities may be compromised. The information derived from the experiments conducted in this study can supplement future work to shape the design and implementation of AR systems within work zone environments, ultimately improving the safety and well-being of workers.

%\section{CRediT Authorship Contribution Statement}
%\textbf{Sepehr Sabeti:} Conceptualization, Methodology, Software, Data curation, Data collection, Data analysis, Writing – original draft, Writing – review \& editing, Visualization.\textbf{Omidreza Shoghli:} Conceptualization, Methodology, Visualization,  Writing – review \& editing, Supervision, Funding acquisition. \textbf{Fatemeh Banani Ardacani:} Data collection, Data curation, Writing – review \& editing. 

% or


\section*{Acknowledgment}
This research was made possible by partial funding from the National Science Foundation under Award Number 1932524. We thank all individuals who participated in our experiments and extend our appreciation to Mr. Amit Kumar, whose invaluable assistance contributed to conducting the experiments.

\ifCLASSOPTIONcaptionsoff
  \newpage
\fi



%\bibliography{references_v1}
\bibliographystyle{IEEEtran}
\bibliography{bare_jrnl.bib}

\begin{thebibliography}{10}
\providecommand{\url}[1]{#1}
\csname url@samestyle\endcsname
\providecommand{\newblock}{\relax}
\providecommand{\bibinfo}[2]{#2}
\providecommand{\BIBentrySTDinterwordspacing}{\spaceskip=0pt\relax}
\providecommand{\BIBentryALTinterwordstretchfactor}{4}
\providecommand{\BIBentryALTinterwordspacing}{\spaceskip=\fontdimen2\font plus
\BIBentryALTinterwordstretchfactor\fontdimen3\font minus \fontdimen4\font\relax}
\providecommand{\BIBforeignlanguage}[2]{{%
\expandafter\ifx\csname l@#1\endcsname\relax
\typeout{** WARNING: IEEEtran.bst: No hyphenation pattern has been}%
\typeout{** loaded for the language `#1'. Using the pattern for}%
\typeout{** the default language instead.}%
\else
\language=\csname l@#1\endcsname
\fi
#2}}
\providecommand{\BIBdecl}{\relax}
\BIBdecl

\bibitem{al2020does}
N.~S.~S. Al-Bdairi, ``Does time of day matter at highway work zone crashes?'' \emph{Journal of safety research}, vol.~73, pp. 47--56, 2020.

\bibitem{hou2020study}
G.~Hou and S.~Chen, ``Study of work zone traffic safety under adverse driving conditions with a microscopic traffic simulation approach,'' \emph{Accident Analysis \& Prevention}, vol. 145, p. 105698, 2020.

\bibitem{CDC}
CDC, ``{Highway Work Zone Safety},'' \url{https://www.cdc.gov/niosh/topics/highwayworkzones/default.html/}, 2019, [Online; accessed 2022].

\bibitem{unite2022}
{Unite the Union}, ``Workplace fatal injuries in great britain, 2022,'' \url{https://resources.unitetheunion.org}, published 6 July 2022.

\bibitem{housebill3684}
\BIBentryALTinterwordspacing
{U.S. Congress}. (2021) {H.R.3684 - Infrastructure Investment and Jobs Act}. [Online]. Available: \url{https://www.congress.gov/bill/117th-congress/house-bill/3684}
\BIBentrySTDinterwordspacing

\bibitem{eu2023}
{European Commission - Mobility and Transport}, ``{EU invests €62 billion in sustainable, safe, and efficient transport infrastructure},'' June 2023, accessed: \today.

\bibitem{nnaji2020improving}
C.~Nnaji, J.~Gambatese, H.~W. Lee, and F.~Zhang, ``Improving construction work zone safety using technology: A systematic review of applicable technologies,'' \emph{Journal of traffic and transportation engineering (English edition)}, vol.~7, no.~1, pp. 61--75, 2020.

\bibitem{sakhakarmi2021tactile}
S.~Sakhakarmi, J.~Park, and A.~Singh, ``Tactile-based wearable system for improved hazard perception of worker and equipment collision,'' \emph{Automation in Construction}, vol. 125, p. 103613, 2021.

\bibitem{national2022use}
H.~Brown and P.~Edara, ``Use of smart work zone technologies for improving work zone safety,'' 2022.

\bibitem{tapco_sonoblaster}
``{Sonoblaster Work Zone Intrusion Alarm and Accessories},'' [Online], available at \url{https://www.tapconet.com/product/sonoblaster-work-zone-intrusion-alarm-and-accessories}.

\bibitem{highway_resource_intellicone}
``{Intellicone Incursion Prevention Warning},'' [Online], available at \url{https://www.highwayresource.co.uk/digital-services/intellicone-incursion-prevention-warning/}.

\bibitem{wu2022real}
S.~Wu, L.~Hou, G.~K. Zhang, and H.~Chen, ``Real-time mixed reality-based visual warning for construction workforce safety,'' \emph{Automation in Construction}, vol. 139, p. 104252, 2022.

\bibitem{ramos2022proposal}
J.~Ramos-Hurtado, F.~Mu{\~n}oz-La~Rivera, J.~Mora-Serrano, A.~Deraemaeker, and I.~Valero, ``Proposal for the deployment of an augmented reality tool for construction safety inspection,'' \emph{Buildings}, vol.~12, no.~4, p. 500, 2022.

\bibitem{li2018critical}
X.~Li, W.~Yi, H.-L. Chi, X.~Wang, and A.~P. Chan, ``A critical review of virtual and augmented reality (vr/ar) applications in construction safety,'' \emph{Automation in Construction}, vol.~86, pp. 150--162, 2018.

\bibitem{gilson2020leveraging}
K.~Gilson, J.~Mallela, P.~M. Goodrum \emph{et~al.}, ``Leveraging augmented reality for highway construction,'' Federal Highway Administration (US), Tech. Rep., 2020.

\bibitem{calvi2020effectiveness}
A.~Calvi, F.~D’Amico, C.~Ferrante, and L.~B. Ciampoli, ``Effectiveness of augmented reality warnings on driving behaviour whilst approaching pedestrian crossings: A driving simulator study,'' \emph{Accident Analysis \& Prevention}, vol. 147, p. 105760, 2020.

\bibitem{matviienko2022scootar}
A.~Matviienko, F.~M{\"u}ller, D.~Sch{\"o}n, R.~Fayard, S.~Abaspur, Y.~Li, and M.~M{\"u}hlh{\"a}user, ``E-scootar: Exploring unimodal warnings for e-scooter riders in augmented reality,'' in \emph{CHI Conference on Human Factors in Computing Systems Extended Abstracts}, 2022, pp. 1--7.

\bibitem{von2020augmentation}
T.~Von~Sawitzky, P.~Wintersberger, A.~L{\"o}cken, A.-K. Frison, and A.~Riener, ``Augmentation concepts with huds for cyclists to improve road safety in shared spaces,'' in \emph{Extended Abstracts of the 2020 CHI Conference on Human Factors in Computing Systems}, 2020, pp. 1--9.

\bibitem{zhang2019crash}
K.~Zhang and M.~Hassan, ``Crash severity analysis of nighttime and daytime highway work zone crashes,'' \emph{PLoS one}, vol.~14, no.~8, p. e0221128, 2019.

\bibitem{awolusi2019active}
I.~Awolusi and E.~D. Marks, ``Active work zone safety: preventing accidents using intrusion sensing technologies,'' \emph{Frontiers in built environment}, vol.~5, p.~21, 2019.

\bibitem{li2023semi}
C.~Li, Z.~Qing, P.~Edara, C.~Sun, B.~Balakrishnan, and Y.~Shang, ``Semi-automatic construction of virtual reality environment for highway work zone training using open-source tools,'' in \emph{2023 IEEE Conference on Virtual Reality and 3D User Interfaces Abstracts and Workshops (VRW)}.\hskip 1em plus 0.5em minus 0.4em\relax IEEE, 2023, pp. 489--492.

\bibitem{ergan2022developing}
S.~Ergan, Z.~Zou, S.~D. Bernardes, F.~Zuo, and K.~Ozbay, ``Developing an integrated platform to enable hardware-in-the-loop for synchronous vr, traffic simulation and sensor interactions,'' \emph{Advanced Engineering Informatics}, vol.~51, p. 101476, 2022.

\bibitem{zou2020integrated}
Z.~Zou, S.~D. Bernardes, A.~Kurkcu, S.~Ergan, K.~Ozbay \emph{et~al.}, ``An integrated approach to capture construction workers’ response towards safety alarms using wearable sensors and virtual reality,'' 2020.

\bibitem{horikawa2022comparing}
R.~Horikawa, M.~Ito, K.~Komiya, T.~Nakajima, and H.~Yamana, ``Comparing augmented reality-based display methods to present guiding information,'' in \emph{2022 IEEE 4th Global Conference on Life Sciences and Technologies (LifeTech)}.\hskip 1em plus 0.5em minus 0.4em\relax IEEE, 2022, pp. 22--25.

\bibitem{zaman2021investigating}
F.~Zaman, W.~Drake, J.~Intriligator, A.~Gardony, M.~Natick, and J.~Rife, ``Investigating a virtual reality based subterranean scenario examining augmented reality implications for military operators,'' in \emph{Proceedings of the Human Factors and Ergonomics Society Annual Meeting}, vol.~65, no.~1.\hskip 1em plus 0.5em minus 0.4em\relax SAGE Publications Sage CA: Los Angeles, CA, 2021, pp. 1129--1133.

\bibitem{merenda2019effects}
C.~Merenda, C.~Suga, J.~Gabbard, and T.~Misu, ``Effects of vehicle simulation visual fidelity on assessing driver performance and behavior,'' in \emph{2019 IEEE Intelligent Vehicles Symposium (IV)}.\hskip 1em plus 0.5em minus 0.4em\relax IEEE, 2019, pp. 1679--1686.

\bibitem{sabeti2021toward}
S.~Sabeti, O.~Shoghli, M.~Baharani, and H.~Tabkhi, ``Toward ai-enabled augmented reality to enhance the safety of highway work zones: Feasibility, requirements, and challenges,'' \emph{Advanced Engineering Informatics}, vol.~50, p. 101429, 2021.

\bibitem{gambatese2017work}
J.~A. Gambatese, H.~W. Lee, C.~A. Nnaji \emph{et~al.}, ``Work zone intrusion alert technologies: Assessment and practical guidance,'' Oregon. Dept. of Transportation. Research Section, Tech. Rep., 2017.

\bibitem{park2019embedded}
J.~Park and S.~Sakhakarmi, ``Embedded safety communication system for robust hazard perception of individuals in work zones,'' 2019.

\bibitem{stout1993maintenance}
D.~Stout, J.~Graham, B.~Bryant-Fields, J.~Migletz, J.~Fish, and F.~Hanscom, ``Maintenance work zone safety devices development and evaluation,'' \emph{Strategic Highway Research Program report SHRP-H-371, National Research Council, Washington DC}, 1993.

\bibitem{eseonu2018reducing}
C.~Eseonu, J.~Gambatese, and C.~Nnaji, ``Reducing highway construction fatalities through improved adoption of safety technologies,'' \emph{The Centre for Construction Research and Training (CPWR Small Study No. 17-4-PS)}, 2018.

\bibitem{marks2017active}
E.~Marks and I.~Vereen, ``Active work zone safety using emerging technologies 2017,'' University Transportation Center for Alabama, Tech. Rep., 2017.

\bibitem{chan2020incorporating}
K.~Chan, J.~Louis, and A.~Albert, ``Incorporating worker awareness in the generation of hazard proximity warnings,'' \emph{Sensors}, vol.~20, no.~3, p. 806, 2020.

\bibitem{kim2023signal}
K.~Kim, I.~Jeong, and Y.~K. Cho, ``Signal processing and alert logic evaluation for iot--based work zone proximity safety system,'' \emph{Journal of Construction Engineering and Management}, vol. 149, no.~2, p. 05022018, 2023.

\bibitem{yun2020multimodal}
H.~Yun and J.~H. Yang, ``Multimodal warning design for take-over request in conditionally automated driving,'' \emph{European transport research review}, vol.~12, pp. 1--11, 2020.

\bibitem{matviienko2018augmenting}
A.~Matviienko, S.~Ananthanarayan, S.~S. Borojeni, Y.~Feld, W.~Heuten, and S.~Boll, ``Augmenting bicycles and helmets with multimodal warnings for children,'' in \emph{Proceedings of the 20th International Conference on Human-Computer Interaction with Mobile Devices and Services}, 2018, pp. 1--13.

\bibitem{wang2022effect}
Y.~Wang, B.~Wu, S.~Ma, D.~Wang, T.~Gan, H.~Liu, and Z.~Yang, ``Effect of mapping characteristic on audiovisual warning: Evidence from a simulated driving study,'' \emph{Applied ergonomics}, vol.~99, p. 103638, 2022.

\bibitem{lazaro2021interaction}
M.~J. Lazaro, S.~Kim, J.~Lee, J.~Chun, and M.-H. Yun, ``Interaction modalities for notification signals in augmented reality,'' in \emph{Proceedings of the 2021 International Conference on Multimodal Interaction}, 2021, pp. 470--477.

\bibitem{wiegand2019incarar}
G.~Wiegand, C.~Mai, K.~Holl{\"a}nder, and H.~Hussmann, ``Incarar: A design space towards 3d augmented reality applications in vehicles,'' in \emph{Proceedings of the 11th international conference on automotive user interfaces and interactive vehicular applications}, 2019, pp. 1--13.

\bibitem{zhou2018arve}
P.~Zhou, W.~Zhang, T.~Braud, P.~Hui, and J.~Kangasharju, ``Arve: Augmented reality applications in vehicle to edge networks,'' in \emph{Proceedings of the 2018 Workshop on Mobile Edge Communications}, 2018, pp. 25--30.

\bibitem{posner2005timing}
M.~I. Posner, ``Timing the brain: Mental chronometry as a tool in neuroscience,'' \emph{PLoS biology}, vol.~3, no.~2, p. e51, 2005.

\bibitem{fernandez2011relation}
J.~Fernandez-Ruiz, W.~Wong, I.~T. Armstrong, and J.~R. Flanagan, ``Relation between reaction time and reach errors during visuomotor adaptation,'' \emph{Behavioural brain research}, vol. 219, no.~1, pp. 8--14, 2011.

\bibitem{li2004transformations}
S.-C. Li, U.~Lindenberger, B.~Hommel, G.~Aschersleben, W.~Prinz, and P.~B. Baltes, ``Transformations in the couplings among intellectual abilities and constituent cognitive processes across the life span,'' \emph{Psychological science}, vol.~15, no.~3, pp. 155--163, 2004.

\bibitem{willoughby2018benefits}
M.~T. Willoughby, C.~B. Blair, L.~J. Kuhn, and B.~E. Magnus, ``The benefits of adding a brief measure of simple reaction time to the assessment of executive function skills in early childhood,'' \emph{Journal of experimental child psychology}, vol. 170, pp. 30--44, 2018.

\bibitem{mueckstein2022modality}
M.~Mueckstein, S.~Heinzel, U.~Granacher, M.~Brahms, M.~A. Rapp, and C.~Stelzel, ``Modality-specific effects of mental fatigue in multitasking,'' \emph{Acta Psychologica}, vol. 230, p. 103766, 2022.

\bibitem{maslovat2019effect}
D.~Maslovat, S.~T. Klapp, C.~J. Forgaard, R.~Chua, and I.~M. Franks, ``The effect of response complexity on simple reaction time occurs even with a highly predictable imperative stimulus,'' \emph{Neuroscience Letters}, vol. 704, pp. 62--66, 2019.

\bibitem{richer2014impact}
N.~Richer, N.~Paquet, and Y.~Lajoie, ``Impact of age and obstacles on navigation precision and reaction time during blind navigation in dual-task conditions,'' \emph{Gait \& posture}, vol.~39, no.~3, pp. 835--840, 2014.

\bibitem{lu2020increased}
K.~Lu, J.~M. Nicholas, S.-N. James, C.~A. Lane, T.~D. Parker, A.~Keshavan, S.~E. Keuss, S.~M. Buchanan, H.~Murray-Smith, D.~M. Cash \emph{et~al.}, ``Increased variability in reaction time is associated with amyloid beta pathology at age 70,'' \emph{Alzheimer's \& Dementia: Diagnosis, Assessment \& Disease Monitoring}, vol.~12, no.~1, p. e12076, 2020.

\bibitem{woods2015age}
D.~L. Woods, J.~M. Wyma, E.~W. Yund, T.~J. Herron, and B.~Reed, ``Age-related slowing of response selection and production in a visual choice reaction time task,'' \emph{Frontiers in human neuroscience}, vol.~9, p. 193, 2015.

\bibitem{greenwald2022sequential}
A.~G. Greenwald and K.~E. Rosenberg, ``Sequential effects of distracting stimuli in a selective attention reaction time task,'' in \emph{Attention and performance VII}.\hskip 1em plus 0.5em minus 0.4em\relax Routledge, 2022, pp. 487--504.

\bibitem{langner2010mental}
R.~Langner, M.~B. Steinborn, A.~Chatterjee, W.~Sturm, and K.~Willmes, ``Mental fatigue and temporal preparation in simple reaction-time performance,'' \emph{Acta psychologica}, vol. 133, no.~1, pp. 64--72, 2010.

\bibitem{maylor1992effects}
E.~A. Maylor, P.~Rabbitt, G.~James, and S.~Kerr, ``Effects of alcohol, practice, and task complexity on reaction time distributions,'' \emph{The Quarterly Journal of Experimental Psychology Section A}, vol.~44, no.~1, pp. 119--139, 1992.

\bibitem{nnaji2020case}
C.~Nnaji, A.~A. Karakhan, J.~Gambatese, and H.~W. Lee, ``Case study to evaluate work-zone safety technologies in highway construction,'' \emph{Practice Periodical on Structural Design and Construction}, vol.~25, no.~3, p. 05020004, 2020.

\bibitem{thapa2021using}
D.~Thapa and S.~Mishra, ``Using worker's naturalistic response to determine and analyze work zone crashes in the presence of work zone intrusion alert systems,'' \emph{Accident Analysis \& Prevention}, vol. 156, p. 106125, 2021.

\bibitem{yang2023vibrotactile}
X.~Yang and N.~Roofigari-Esfahan, ``Vibrotactile alerting to prevent accidents in highway construction work zones: An exploratory study,'' \emph{Sensors}, vol.~23, no.~12, p. 5651, 2023.

\bibitem{bottani2019augmented}
E.~Bottani and G.~Vignali, ``Augmented reality technology in the manufacturing industry: A review of the last decade,'' \emph{IISE Transactions}, vol.~51, no.~3, pp. 284--310, 2019.

\bibitem{kelly2018arcadia}
A.~Kelly, R.~B. Shapiro, J.~de~Halleux, and T.~Ball, ``Arcadia: A rapid prototyping platform for real-time tangible interfaces,'' in \emph{Proceedings of the 2018 CHI Conference on Human Factors in Computing Systems}, 2018, pp. 1--8.

\bibitem{muller2021spatialproto}
L.~M{\"u}ller, K.~Pfeuffer, J.~Gugenheimer, B.~Pfleging, S.~Prange, and F.~Alt, ``Spatialproto: Exploring real-world motion captures for rapid prototyping of interactive mixed reality,'' in \emph{Proceedings of the 2021 CHI Conference on Human Factors in Computing Systems}, 2021, pp. 1--13.

\bibitem{hettig2018ar}
J.~Hettig, S.~Engelhardt, C.~Hansen, and G.~Mistelbauer, ``Ar in vr: assessing surgical augmented reality visualizations in a steerable virtual reality environment,'' \emph{International journal of computer assisted radiology and surgery}, vol.~13, no.~11, pp. 1717--1725, 2018.

\bibitem{terrier2018evaluation}
R.~Terrier, F.~Argelaguet, J.-M. Normand, and M.~Marchal, ``Evaluation of ar inconsistencies on ar placement tasks: A vr simulation study,'' in \emph{International Conference on Virtual Reality and Augmented Reality}.\hskip 1em plus 0.5em minus 0.4em\relax Springer, 2018, pp. 190--210.

\bibitem{grandi2021design}
J.~G. Grandi, Z.~Cao, M.~Ogren, and R.~Kopper, ``Design and simulation of next-generation augmented reality user interfaces in virtual reality,'' in \emph{2021 IEEE Conference on Virtual Reality and 3D User Interfaces Abstracts and Workshops (VRW)}.\hskip 1em plus 0.5em minus 0.4em\relax IEEE, 2021, pp. 23--29.

\bibitem{burova2020utilizing}
A.~Burova, J.~M{\"a}kel{\"a}, J.~Hakulinen, T.~Keskinen, H.~Heinonen, S.~Siltanen, and M.~Turunen, ``Utilizing vr and gaze tracking to develop ar solutions for industrial maintenance,'' in \emph{Proceedings of the 2020 CHI Conference on Human Factors in Computing Systems}, 2020, pp. 1--13.

\bibitem{doi:10.1080/10803548.2023.2295660}
\BIBentryALTinterwordspacing
S.~Sabeti, N.~Morris, and O.~Shoghli, ``Mixed-method usability investigation of arrows: augmented reality for roadway work zone safety,'' \emph{International Journal of Occupational Safety and Ergonomics}, vol.~30, no.~1, pp. 292--303, 2024, pMID: 38097505. [Online]. Available: \url{https://doi.org/10.1080/10803548.2023.2295660}
\BIBentrySTDinterwordspacing

\bibitem{mutcd2006manual}
T.~MUTCD, ``Manual on uniform traffic control devices,'' \emph{Texas Department of Transportation, Austin}, 2006.

\bibitem{tizen}
``{Tizen IoT Documentation},'' \url{https://docs.tizen.org/iot/api/5.0/tizen-iot-headed/group__CAPI__SYSTEM__FEEDBACK__MODULE.html}, accessed: [\today].

\bibitem{goenarjo2020cerebral}
R.~Goenarjo, L.~Bosquet, N.~Berryman, V.~Metier, A.~Perrochon, S.~A. Fraser, and O.~Dupuy, ``Cerebral oxygenation reserve: The relationship between physical activity level and the cognitive load during a stroop task in healthy young males,'' \emph{International Journal of Environmental Research and Public Health}, vol.~17, no.~4, p. 1406, 2020.

\bibitem{snyder2022aperture}
N.~A. Snyder and M.~E. Cinelli, ``Aperture crossing in virtual reality: Physical fatigue delays response time,'' \emph{Journal of Motor Behavior}, vol.~54, no.~4, pp. 429--437, 2022.

\bibitem{sabeti2022toward}
S.~Sabeti, O.~Shoghli, and H.~Tabkhi, ``Toward wi-fi-enabled real-time communication for proactive safety systems in highway work zones: A case study,'' in \emph{Construction Research Congress 2022}, 2022, pp. 1166--1173.

\bibitem{jain2015comparative}
A.~Jain, R.~Bansal, A.~Kumar, and K.~Singh, ``A comparative study of visual and auditory reaction times on the basis of gender and physical activity levels of medical first year students,'' \emph{International Journal of Applied and Basic Medical Research}, vol.~5, no.~2, p. 124, 2015.

\bibitem{googlemlkit}
``Ml kit pose detection api,'' \url{https://developers.google.com/ml-kit/vision/pose-detection}, accessed on: [Insert Date].

\bibitem{legg2021does}
H.~S. Legg, C.~M. Arnold, C.~Trask, and J.~L. Lanovaz, ``Does functional performance and upper body strength predict upper extremity reaction and movement time in older women?'' \emph{Human Movement Science}, vol.~77, p. 102796, 2021.

\bibitem{godfrey2008direct}
A.~Godfrey, R.~Conway, D.~Meagher, and G.~{\'O}Laighin, ``Direct measurement of human movement by accelerometry,'' \emph{Medical engineering \& physics}, vol.~30, no.~10, pp. 1364--1386, 2008.

\bibitem{uddin2020body}
M.~Z. Uddin, M.~M. Hassan, A.~Alsanad, and C.~Savaglio, ``A body sensor data fusion and deep recurrent neural network-based behavior recognition approach for robust healthcare,'' \emph{Information Fusion}, vol.~55, pp. 105--115, 2020.

\bibitem{begg2005machine}
R.~Begg and J.~Kamruzzaman, ``A machine learning approach for automated recognition of movement patterns using basic, kinetic and kinematic gait data,'' \emph{Journal of biomechanics}, vol.~38, no.~3, pp. 401--408, 2005.

\bibitem{sato2006wavelet}
H.~Sato, N.~Tanaka, M.~Uchida, Y.~Hirabayashi, M.~Kanai, T.~Ashida, I.~Konishi, and A.~Maki, ``Wavelet analysis for detecting body-movement artifacts in optical topography signals,'' \emph{Neuroimage}, vol.~33, no.~2, pp. 580--587, 2006.

\bibitem{jia2019effect}
W.~Jia and L.~Shi, ``The effect of auditory stimuli in audio-visual two-source integration,'' in \emph{2019 11th International Conference on Intelligent Human-Machine Systems and Cybernetics (IHMSC)}, vol.~1.\hskip 1em plus 0.5em minus 0.4em\relax IEEE, 2019, pp. 145--148.

\bibitem{geitner2019comparison}
C.~Geitner, F.~Biondi, L.~Skrypchuk, P.~Jennings, and S.~Birrell, ``The comparison of auditory, tactile, and multimodal warnings for the effective communication of unexpected events during an automated driving scenario,'' \emph{Transportation research part F: traffic psychology and behaviour}, vol.~65, pp. 23--33, 2019.

\bibitem{lee2023investigating}
S.~Lee, J.~Hong, G.~Jeon, J.~Jo, S.~Boo, H.~Kim, S.~Jung, J.~Park, I.~Choi, and S.~Kim, ``Investigating effects of multimodal explanations using multiple in-vehicle displays for takeover request in conditionally automated driving,'' \emph{Transportation Research Part F: Traffic Psychology and Behaviour}, vol.~96, pp. 1--22, 2023.

\bibitem{douce2020sensory}
L.~Douc{\'e} and C.~Adams, ``Sensory overload in a shopping environment: Not every sensory modality leads to too much stimulation,'' \emph{Journal of Retailing and Consumer Services}, vol.~57, p. 102154, 2020.

\bibitem{katic2015system}
D.~Kati{\'c}, P.~Spengler, S.~Bodenstedt, G.~Castrillon-Oberndorfer, R.~Seeberger, J.~Hoffmann, R.~Dillmann, and S.~Speidel, ``A system for context-aware intraoperative augmented reality in dental implant surgery,'' \emph{International journal of computer assisted radiology and surgery}, vol.~10, pp. 101--108, 2015.

\bibitem{wueffect}
R.~Wu and H.-T. Chen, ``The effect of visual and auditory modality mismatching between distraction and warning on pedestrian street crossing behavior.''

\bibitem{edet2022evaluation}
U.~Edet and D.~D. Mann, ``Evaluation of warning methods for remotely supervised autonomous agricultural machines,'' \emph{Journal of agricultural safety and health}, vol.~28, no.~1, pp. 1--17, 2022.

\bibitem{williams2014effects}
T.~Williams, T.~Esposito, S.~Hu, D.~Mahoney, and K.~Paulson, ``Effects of varying audio frequencies on reaction time and muscular activity,'' 2014.

\bibitem{navarro2023mathematical}
E.~A. Navarro and E.~Navarro-Modesto, ``A mathematical model and experimental procedure to analyze the cognitive effects of audio frequency magnetic fields,'' \emph{Frontiers in Human Neuroscience}, vol.~17, p. 1135511, 2023.

\bibitem{fyhrie2016work}
P.~B. Fyhrie, A.~H. Maintenance, and C.~T. R.~C. (Calif.), \emph{Work Zone Intrusion Alarms for Highway Workers}.\hskip 1em plus 0.5em minus 0.4em\relax Caltrans Division of Research, Innovation and System Information, 2016.

\bibitem{mosunmola2022countermeasures}
O.~Mosunmola~Aroke, I.~Sylvester~Onuchukwu, B.~Esmaeili, and A.~M. Flintsch, ``Countermeasures to reduce truck-mounted attenuator (tma) crashes: a state-of-the-art review,'' \emph{Future transportation}, vol.~2, no.~2, pp. 425--452, 2022.

\bibitem{nnaji2018developing}
C.~Nnaji, A.~Lee, and J.~Gambatese, ``Developing a decision-making framework to select safety technologies for highway construction,'' \emph{Journal of Construction Engineering and Management}, vol. 144, no.~4, p. 04018016, 2018.

\bibitem{hasan2021distributed}
M.~Hasan, D.~Perez, Y.~Shen, and H.~Yang, ``Distributed microscopic traffic simulation with human-in-the-loop enabled by virtual reality technologies,'' \emph{Advances in Engineering Software}, vol. 154, p. 102985, 2021.

\bibitem{harris2021exploring}
D.~J. Harris, G.~Buckingham, M.~R. Wilson, J.~Brookes, F.~Mushtaq, M.~Mon-Williams, and S.~J. Vine, ``Exploring sensorimotor performance and user experience within a virtual reality golf putting simulator,'' \emph{Virtual Reality}, vol.~25, pp. 647--654, 2021.

\bibitem{mills2020virtual}
B.~Mills, P.~Dykstra, S.~Hansen, A.~Miles, T.~Rankin, L.~Hopper, L.~Brook, and D.~Bartlett, ``Virtual reality triage training can provide comparable simulation efficacy for paramedicine students compared to live simulation-based scenarios,'' \emph{Prehospital Emergency Care}, vol.~24, no.~4, pp. 525--536, 2020.

\bibitem{hoeg2017binaural}
E.~R. Hoeg, L.~J. Gerry, L.~Thomsen, N.~C. Nilsson, and S.~Serafin, ``Binaural sound reduces reaction time in a virtual reality search task,'' in \emph{2017 IEEE 3rd VR workshop on sonic interactions for virtual environments (SIVE)}.\hskip 1em plus 0.5em minus 0.4em\relax IEEE, 2017, pp. 1--4.

\bibitem{fang2020computer}
W.~Fang, L.~Ding, P.~E. Love, H.~Luo, H.~Li, F.~Pena-Mora, B.~Zhong, and C.~Zhou, ``Computer vision applications in construction safety assurance,'' \emph{Automation in Construction}, vol. 110, p. 103013, 2020.

\bibitem{khan2022fall}
M.~Khan, R.~Khalid, S.~Anjum, S.~V.-T. Tran, and C.~Park, ``Fall prevention from scaffolding using computer vision and iot-based monitoring,'' \emph{Journal of Construction Engineering and Management}, vol. 148, no.~7, p. 04022051, 2022.

\bibitem{guo2021computer}
B.~H. Guo, Y.~Zou, Y.~Fang, Y.~M. Goh, and P.~X. Zou, ``Computer vision technologies for safety science and management in construction: A critical review and future research directions,'' \emph{Safety science}, vol. 135, p. 105130, 2021.

\end{thebibliography}
\IEEEtriggercmd{\enlargethispage{-5in}}
\IEEEtriggeratref{1}  
\begin{IEEEbiography}[{\includegraphics[width=1in,height=1.25in,clip,keepaspectratio]{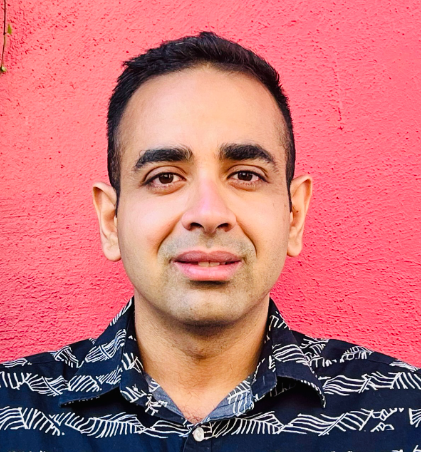}}]{Sepehr Sabeti}
received his Ph.D. degree from the University of North Carolina at Charlotte in 2023, under the supervision of Dr. Omidreza Shoghli. He is currently a Simulation Engineer at Leidos. His research interests focus on AI-driven predictive analytics, Machine Learning, real-time Computer Vision for robotics applications, and data-driven human performance measurement within Augmented Reality and Virtual Reality environments. 
\end{IEEEbiography}
\begin{IEEEbiography}[{\includegraphics[width=1in,height=1.25in,clip,keepaspectratio]{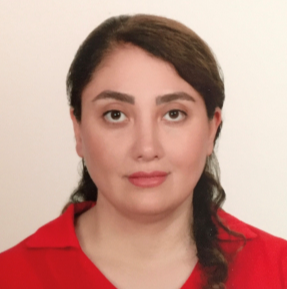}}]{Fatemeh Banani Ardacani}
received her Master's degree in Civil Engineering from Amirkabir University in 2017. She is currently a Ph.D. candidate in civil engineering with a concentration in transportation at the University of North Carolina at Charlotte. Her research is focused on the relationship between brain behavior and human sensing, particularly as it pertains to vulnerable road users and roadway work zone safety. She is also interested in the application of virtual reality for human sensing in transportation.

\end{IEEEbiography}
\begin{IEEEbiography}[{\includegraphics[width=1in,height=1.25in,clip,keepaspectratio]{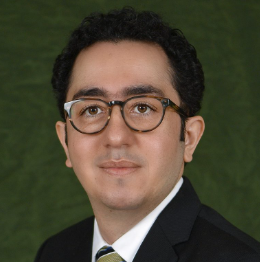}}]{Omidreza Shoghli}
received his Ph.D. in Civil Engineering from Virginia Tech in 2014. Currently he is an Associate Professor and the Director of the Smart Infrastructure Asset Management (SIAM) Lab at the University of North Carolina at Charlotte. His research interests lie at the intersection of human factors and the safety of vulnerable road users, employing cutting-edge Virtual Reality and Augmented Reality technologies. He is also interested in AI-based predictive analytics and decision optimization.
\end{IEEEbiography}
\end{document}